\documentclass[preprint,12pt]{elsarticle}

\usepackage{lineno,hyperref}
\usepackage{hyperref}
\usepackage{graphicx}
\usepackage{amssymb}
\usepackage{amsmath}
\usepackage[dvipsnames]{xcolor}
\usepackage{ctable}
\usepackage{url}
\usepackage{bm}

\newcommand\Tstrut{\rule{0pt}{3.5ex}}       % "top" strut
\newcommand\Bstrut{\rule[-2.5ex]{0pt}{0pt}} % "bottom" strut
\newcommand{\TBstrut}{\Tstrut\Bstrut} % top&bottom struts

\journal{Journal of Informetrics}

\biboptions{authoryear}
\begin{document}

\begin{frontmatter}

\title{Predicting the Citation Count and CiteScore of Journals One Year in Advance\tnoteref{t1}}

\tnotetext[t1]{Funding: This work was supported by the Natural Sciences and Engineering Research Council of Canada (NSERC) [grant number RGPIN-2016-06253].}

\author[1]{William L. Croft\corref{mycorrespondingauthor}}
\ead{leecroft@cmail.carleton.ca}
\cortext[mycorrespondingauthor]{Corresponding author}

\author[1]{J{\"o}rg-R{\"u}diger Sack}
\ead{sack@scs.carleton.ca}

\address[1]{School of Computer Science, Carleton University. Ottawa, Canada.}

\begin{abstract}
Prediction of the future performance of academic journals is a task that can benefit a variety of stakeholders including editorial staff, publishers, indexing services, researchers, university administrators and granting agencies. Using historical data on journal performance, this can be framed as a machine learning regression problem. In this work, we study two such regression tasks: 1) prediction of the number of citations a journal will receive during the next calendar year, and 2) prediction of the Elsevier CiteScore a journal will be assigned for the next calendar year. To address these tasks, we first create a dataset of historical bibliometric data for journals indexed in Scopus. We propose the use of neural network models trained on our dataset to predict the future performance of journals. To this end, we perform feature selection and model configuration for a Multi-Layer Perceptron and a Long Short-Term Memory. Through experimental comparisons to heuristic prediction baselines and classical machine learning models, we demonstrate superior performance in our proposed models for the prediction of future citation and CiteScore values.
\end{abstract}


\end{frontmatter}

\section{Introduction}

Academic journals are one of the primary channels through which scientific research is published and disseminated to both research communities and wider audiences. The performance of such journals is a matter of interest to a variety of stakeholders for diverse reasons. While information on the past and present performance of journals is crucial in this regard, the ability to predict future performance can also be highly valuable in the derivation of complementary information as it lends itself to a deeper understanding of the state of the journals. The more detail that can be provided on the probable future performance of a journal, the easier it becomes to make decisions that are impacted by the future state of the journal. There are numerous situations in which this type of information is of value:

\begin{itemize}
	
	\item Editorial boards for journals are directly responsible for ensuring a consistent level of quality in the journals they manage. Decisions around journal management and editorial board composition are made with the future performance of the journal explicitly in mind.
	
	\item Indexing services must ensure that only journals of sufficient quality are accepted for indexing. Knowledge about the future performance of journals therefore helps to inform decision of whether new journals should be accepted and whether currently indexed journals should be removed.
	
	\item Publishers must monitor the performance of journals (both their own and those of other publishers) to inform strategic decision-making. Knowledge about the trajectory of journal performance and its projection into the future can help to make decisions on the acquisition of existing journals and the launching of new journals.
	
	\item Some granting agencies maintain lists of journals categorized by strength. These lists are updated regularly and early indications of which journals to move up or down might be helpful for granting agencies and applicants.
	
	\item To academic institutions and countries, the quality of relevant journals may serve as an indicator in the assessment of their scientific output. Information on the expected performance of journals may help to inform decisions on budget allocation.
	
	\item For research groups and authors, information on journal quality is often factored into the decision on where to submit their work. Due to the lengthy duration of most review and publication processes, the future performance of a journal is highly relevant in making this decision.
	
	\item To librarians building collections of journals, the performance of a journal may help to assess its merit for selection. Information on future performance helps in the selection of journals leading to a sustainable collection.
	
\end{itemize}

The information needed to quantitatively measure the performance of a journal for a future point in time is necessarily incomplete since the events contributing to any relevant performance metrics have not yet fully occurred. It is therefore necessary to use predictive modeling to produce this information. In this paper, we study the task of predicting the future performance of journals. To this end, we construct a dataset on the historical performance of journals and develop predictive models trained on this data.

\subsection{Problem Domain}

We consider two predictive tasks for the purpose of projecting the future performance of journals.

\medskip
\noindent \textbf{Task 1:} For a given journal, predict how many citations it will receive in the next calendar year.
\medskip

Alone, the number of citations received may not be indicative of journal performance. However, this value offers great flexibility in its usage as it can be fed into various calculations or models to assess journal performance. Simple examples include examination of the predicted value in relation to the average number of yearly publications of the journal, or plotting the citations received by the journal during past years with the predicted value used to extrapolate the plotted series. The number of citations may also be used as an input feature to more sophisticated models that examine multiple aspects of a journal.

\medskip
\noindent \textbf{Task 2:} For a given journal, predict its Elsevier CiteScore \cite{25} for the next calendar year.
\medskip

As an impact-based metric, CiteScore values are useful indicators of the performance of journals. We have chosen CiteScore over other impact-based measures such as Journal Impact Factor due to its transparent and reproducible calculation process. Further information on the strengths and weaknesses of various measures of impact is provided in Section \ref{sec:bibliometrics}.

While journal performance often does not change greatly from one year to the next, there is still ample room for improvement over simple heuristic predictions (e.g., predicting the same value for the next year or a value following from a linear trend), as later demonstrated in our experimental comparisons (see Sections \ref{sec:experiments} and \ref{sec:lstm_analysis}). Furthermore, the task of predicting values one year in advance is an important step towards making longer range predictions.

We formalize the two predictive tasks as regression problems. For each, we consider the use of various input features on the historical performance of journals to predict the desired value. We investigate the potential of deep learning approaches in this context and propose two such models which achieve good performance.

\subsection{Contributions and Paper Outline}

The primary contributions of this work are as follows:

\begin{itemize}
	\item We have collected and formed a dataset covering publication metrics for over 24,000 journals indexed in Scopus during the period of 2000 through 2020 (Section \ref{sec:dataset}).
	
	\item We have applied four standard machine learning algorithms (Linear Regression, Decision Tree, Random Forest and k-Nearest Neighbors) to establish baselines for the predictive tasks we have set out (Subections \ref{sec:standard_models} and \ref{sec:model_selection}).
	
	\item We have identified and configured two deep neural network models appropriate for our predictive tasks, one of which is particularly well-suited to handling the type of timeseries data we have compiled (Subections \ref{sec:neural_net_models} and \ref{sec:model_selection}).
	
	\item We have conducted experimental evaluations of our proposed models, demonstrating improvements over the selected baselines (Section \ref{sec:experiments}).
\end{itemize}

We begin by providing a review of relevant literature in Section \ref{sec:lit_rev}. In Section \ref{sec:preliminaries}, we provide preliminary background information on concepts used throughout the paper. We describe the data that we have collected and the process we have applied to build a dataset from it in Section \ref{sec:dataset}. We then describe the features we have selected for the predictive tasks and the machine learning models we have identified to handle these tasks in Section \ref{sec:models}. We provide details on the experiments we have performed for the configuration of the models in Section \ref{sec:model_selection}. In Section \ref{sec:experiments}, we detail the experimental comparisons we have conducted on finalized configurations of the predictive models and the results we have obtained. We conclude the paper in Section \ref{sec:conclusion}.

\section{Literature Review} \label{sec:lit_rev}

The evaluation of performance in academic journals is dependent upon the definition of metrics that provide meaningful information on various properties of journals. Among such metrics, those which examine the impact of journals are often associated with some indication of the quality of journals. We provide here a review of some of the most commonly-used measures of impact and the works that have designed predictive tasks around them. We additionally provide a brief review of predictive tasks related to paper-level measures of impact.

\subsection{Journal Bibliometrics} \label{sec:bibliometrics}

The field of bibliometrics pertains to the statistical analysis of publications and their channels of publication (i.e., sources such as journals) \cite{7,4,3}. In the context of academic journals, bibliometrics often focuses on measures built around numbers of citations and publications to assess the quality of journals. While general terms such as performance or quality are often used when describing the goal of journal-centric bibliometrics, these notions are typically vague and ill-defined. The more precise intent of many bibliometric measures is to capture scientific impact, generally employing some form of a ratio of citations received over documents published \cite{1}. The notion of an ``impact factor" in scientific literature was first introduced in the context of the proposal by Garfield to develop a citation index to track the linkages between indexed publications and their citations \cite{12}. Impact factor refers to the effect a publication or a corpus of published literature has on the research community, observed through the lens of citations attributed to the published work. Although the measure of impact has a precise meaning that was not originally intended to be employed as a general reflection of quality, it has found widespread usage as a proxy for journal quality \cite{2}.

Journal Impact Factor (JIF) \cite{9} was the first proposed measure of journal impact. It is measured as the number of citations received during year X on all publications of a journal from years X-1 and X-2 divided by the number of substantive articles and reviews released during the same two years. This offers a view of the average impact of the journal publications, restricted to a two year window in order to focus on recent material for greater relevance as a year-to-year metric. All values are calculated using the Science Citation Index (SCI), an indexing database of academic literature that contains only material from journals deemed to be of sufficiently high quality \cite{9,26}. JIF has enjoyed widespread usage and popularity, largely due to a broad coverage of journals and a simple yet meaningful definition \cite{6}. Despite this, it has also been subject to a great deal of criticism. Differences between the publication types used in the numerator and denominator of the JIF have led to concerns over the appropriateness of the metric for journals with large disparities between published documents and citeable documents, and by extension, concerns over the potential for exploitation of JIF values by selective publishing of certain document types \cite{5}. The metric has also been criticized for its opaque calculation (as the contents of the SCI database are not publicly accessible) and for its short window of time, which gives preference to fields of research that have faster turnaround times for the generation of citations on published material \cite{5}.

As an alternative to JIF, Elsevier launched CiteScore in 2016 to provide a competing measure of impact that addresses some of the shortcomings of JIF \cite{24}. Perhaps chief among these is a transparent calculation process \cite{10,11}; the formula is public and the publication data fed into the formula can be accessed by Scopus users. Scopus is a curated database of journals selected by an independent board using 14 selection criteria grouped into 5 categories\footnote{See e.g., \url{https://iro.bu.edu.eg/assets/img/workshop/1kh36e33/Scopus\%20Indexing\%20Criteria_v2.pdf}}. Furthermore, the same document types are used in both the numerator and denominator of the CiteScore formula in order to mitigate some of the potential for manipulation of the measure \cite{8,10}. The original CiteScore calculation used a formula similar to JIF extended to a three year window. In 2020, this was further extended to a four year window with a notable change to the numerator such that it includes citations received over the duration of the window as opposed to the year after the window \cite{25}. These adjustments are intended to provide a more robust metric that is fair to a variety of research fields that may differ in the rates at which publications accrue citations \cite{27}.

To address other criticisms of existing impact-based measures a variety of other measures have been proposed. Due to differences in citation behaviours across research fields, it is well understood that journals from different fields may have vastly different distributions of impact \cite{1,7}. To address this concern, Source Normalized Impact per Paper (SNIP) was proposed as a measure that applies normalization across research fields to make the results more comparable across different fields \cite{15}. Other concerns have been raised over relative importance of citations from different sources, leading to the design of prestige-based measures \cite{5}. Measures such as Eigenfactor Score \cite{13} and SCImago Journal Rank (SJR) \cite{14} assign weights to citations based on the impact of the journal from which the citation originated. While these measures offer useful information, they employ complex algorithms and in some cases are not reproducible using public-facing data, leading to criticisms of more difficult interpretability \cite{3,5}. As a result, these measures are often recommended for use in complement to, as opposed to in replacement of, standard impact-based measures \cite{3,4}.

While the information provided here covers many of the journal impact measures that have seen the most usage in practice, it is by no means a comprehensive review. For further details on the topic, we refer to the reader to survey material \cite{4,5}.

\subsection{Journal Predictions}

Given the prevalence of journal bibliometrics and the ongoing indexing scientific publications and citations in large databases, there has been increasing attention paid to the idea of applying data science approaches to this corpus of data. Of particular interest have been predictive tasks focused around the regression of bibliometric values and the classification of journals.

Prediction of JIF in various settings is one of the most commonly approached tasks in this context. Bibliometric features have been studied for correlations with JIF in order to determine their usefulness for JIF regression \cite{18}. Although the study was designed for the regression of JIF values for a completed calendar year from which other bibliometric data is available as input features, it was shown that certain features have positive correlations with JIF and are useful when applied to a linear regression model. Another study has investigated the potential to estimate JIF values using machine learning approaches applied to bibliometric data from open-access databases \cite{20}. Such methods are motivated by the fact that for journals not indexed in SCI, the lack of public accessibility of the indexing database makes it impossible to calculate a journal's hypothetical JIF exactly as it would officially be done. While these approaches are targeted at instances where complete data for the relevant years is available, other studies have aimed to predict JIF in advance of its official release. To reduce the multi-month delay between the end of a calendar year and the release of JIFs, an approach has been proposed to use data from the Web of Science (WoS), a subscription-based access database, to calculate approximate JIFs multiple months in advance \cite{17}. To predict JIFs prior to the end of the year, when publication and citation activities have not yet finished, methods have been proposed to make heuristic extrapolations on the incomplete data in order to derive predictions \cite{16}.

In the context of classification-based predictions, most tasks focus on journal rankings. The ability to automate the ranking of journals is considered of value both to save on costs as well as to potentially mitigate human bias \cite{21,23}. Machine learning classifiers have been applied to categorize journals based on their subject and quartiles \cite{22} and into coarse ranks derived from bibliometric-based ranking systems \cite{19}. Classifiers have also been explored for the automation of ranking that would normally be performed by a panel of experts \cite{21}. Alongside the goal of automated ranking, classification has also been applied in the context of detection of predatory journals \cite{23}.

\subsection{Paper-Level Predictions}

A related area of study is the prediction of impact for individual papers, which is typically embodied by the task of predicting the number of citations a paper will receive. One of the first notable attempts involved modeling a probability distribution for citations received by paper over time in order to develop a heuristic equation to predict the number of citations received based on various paper-specific input parameters \cite{28}. Many subsequent approaches have been taken by treating the task as a regression problem and applying various machine learning models including linear regression \cite{29}, multi-layer perceptrons \cite{34}, convolutional neural networks \cite{32} and recurrent neural networks \cite{33}. Studies have also been conducted to investigate the predictive powers of altmetrics \cite{30} and conference-specific features \cite{31} (in the context of conference papers) for the prediction of citation counts.

For a more detailed review of existing work on paper-level citation prediction and other related tasks, we refer the reader to a survey on the topic \cite{35}.

\section{Preliminaries} \label{sec:preliminaries}

In this section, we provide details on key concepts that are employed throughout the paper. We first define the CiteScore metric with its inputs broken down to the granularity we have applied for the input features of our predictive models. We then provide background details on the deep neural network models that we employ.

\subsection{Elesvier CiteScore}

CiteScore is a journal impact metric reflective of a ratio of the number of citations received over the number of papers published during a specific window of time. The formulation of the metric was updated in 2020 and all usage in the present paper is based on the updated version. More precisely, the metric examines a four year window and is calculated as the number of citations received during the four year window for articles, conference papers, reviews and book chapters published during the same window divided by the number of such documents. The metric calculation is given in Formula \ref{eq:citescore}, where $c_{j,i}$ is the number of citations received during year $j$ for publications from year $i$ and $p_i$ is the number of publications from year $i$.

\begin{equation} \label{eq:citescore}
	cs_y = \frac{\sum\limits_{i=y-3}^y \sum\limits_{j=i}^y c_{j,i}}
	{\sum\limits_{i=y-3}^y p_i}
\end{equation}

\subsection{Neural Network Models}

We have identified two neural network models appropriate for the predictive tasks addressed in our work. The first, a Multi-Layer Perceptron (MLP) \cite{36}, is a feed-forward neural network consisting of one or more hidden layers, each of which is composed of a fully-connected layer followed by a non-linear activation function. In the context of regression, the output layer of the network is itself a fully-connected layer with a number of neurons equal to the desired number of outputs from the model. For both regression tasks defined in our work, the output size will always be 1, giving the predicted value for the next year.

The second neural network model we employ, a Long Short-Term Memory (LSTM) \cite{37}, is a type of Recurrent Neural Network (RNN) \cite{38}. RNNs are designed to process sequential data using one or more recurrent layers. A recurrent layer maintains a hidden state that tracks information about past elements of the input sequence. This information assists in making predictions that are dependent upon patterns in the sequence of input. LSTMs build upon the concept of RNNs by implementing a recurrent layer architecture with gated logic to better manage the preservation of relevant past information. This architecture, in combination with a constant error carousel, helps to deal with the issue of vanishing gradients that hampers performance in standard RNNs.

\section{Journal Performance Dataset} \label{sec:dataset}

In order to train effective instances of machine learning models, particularly when using neural networks, a large amount of training data is required. To our knowledge, there are no publicly available large-scale datasets on the performance of academic journals. However, academic journals are thoroughly indexed by a number of service providers. The relevant data is thus accessible but must be extracted and compiled into a useful dataset.

We have made use of Elsevier's APIs \footnote{\url{https://dev.elsevier.com/api_docs.html}} to programmatically query their indexing services and compile a journal performance dataset. Throughout this section, we describe the data we have extracted, the processing we have applied, and the resultant dataset we have constructed.

\subsection{Data Collection}

To build our dataset, we collect data on annual numbers of publications and citations, as well as the annual SNIP and SJR values for all journals indexed in Elsevier's Scopus database. We focus our attention on metrics derived from these core measures and do not consider altmetrics (e.g., numbers of mentions on social media or in mainstream media) in our present work.

We begin by using the Serial Title Metadata API to gather the IDs of all indexed journals along with their annual SNIP and SJR values. This results in a set of 39,140 journals. Next, we gather the IDs of the publications of each journal using the Scopus Search API. We restrict the gathered publication document types to articles, conference papers, reviews and book chapters to match the CiteScore methodology. Using the counts of the retrieved IDs, we calculate and store the annual numbers of publications for each of the journals. Finally, we use the Scopus Citations Overview API to query the number of citations received by each publication, broken down by the year during which they were received. By calculating the sum of citations received during each year over all publications of a journal, we determine the annual numbers of citations received by the journals and store this data. We additionally calculate and store the annual number of publications that received no citations (counting only from publications released during that year).

To produce a dataset with useful features for CiteScore predictions, we also require information on the number of citations received by a journal during a given year $x$ for publications from another year $y$. Since the citation data retrieved from the API is at the publication-level and is broken down by the year during which the citations were received, we are able to calculate this value. This is achieved by taking the sum of the citations received during year $x$ for all publications of the journal with a publication year of $y$. We calculate and store this information for all valid pairs of years $x$ and $y$ for each journal. Using this along with the annual numbers of publications, we then calculate the annual CiteScore values for each journal.

It is important to note that official CiteScore calculations are made on a yearly basis when data entry from the previous calendar year has been finalized. The calculated values are then frozen going forward. This implies that should any further changes occur to the input values used to calculate CiteScore (e.g., corrections to citation counts during data quality reviews or updates sent from publishers), the official CiteScore values will no longer be consistent with values calculated from the most up-to-date data. We have observed such inconsistencies when comparing our own CiteScore calculations with their official indexed counterparts. Although the differences are generally small, we use our own calculations rather than the official versions in order to preserve the correct relationships between the input values and the CiteScore values in our dataset.

\subsection{Dataset Creation}

From the data we have collected, we construct a dataset of sequences of historical data, at the annual granularity, for each journal. Since we are interested in training models to predict future values from these sequences, we limit the dataset to relatively recent data in order to ensure that the models learn the current trends in publications and citations. To this end, we select only journals that are still active and indexed in 2020 and limit the historical data to go back no farther than 2000. We consider a journal to be active during a given year if it is recorded as having at least one publication during that year. Under this restriction, we have a total of 24,557 active journals.

\begin{figure}
	\centering
	\includegraphics[width=0.7\textwidth]{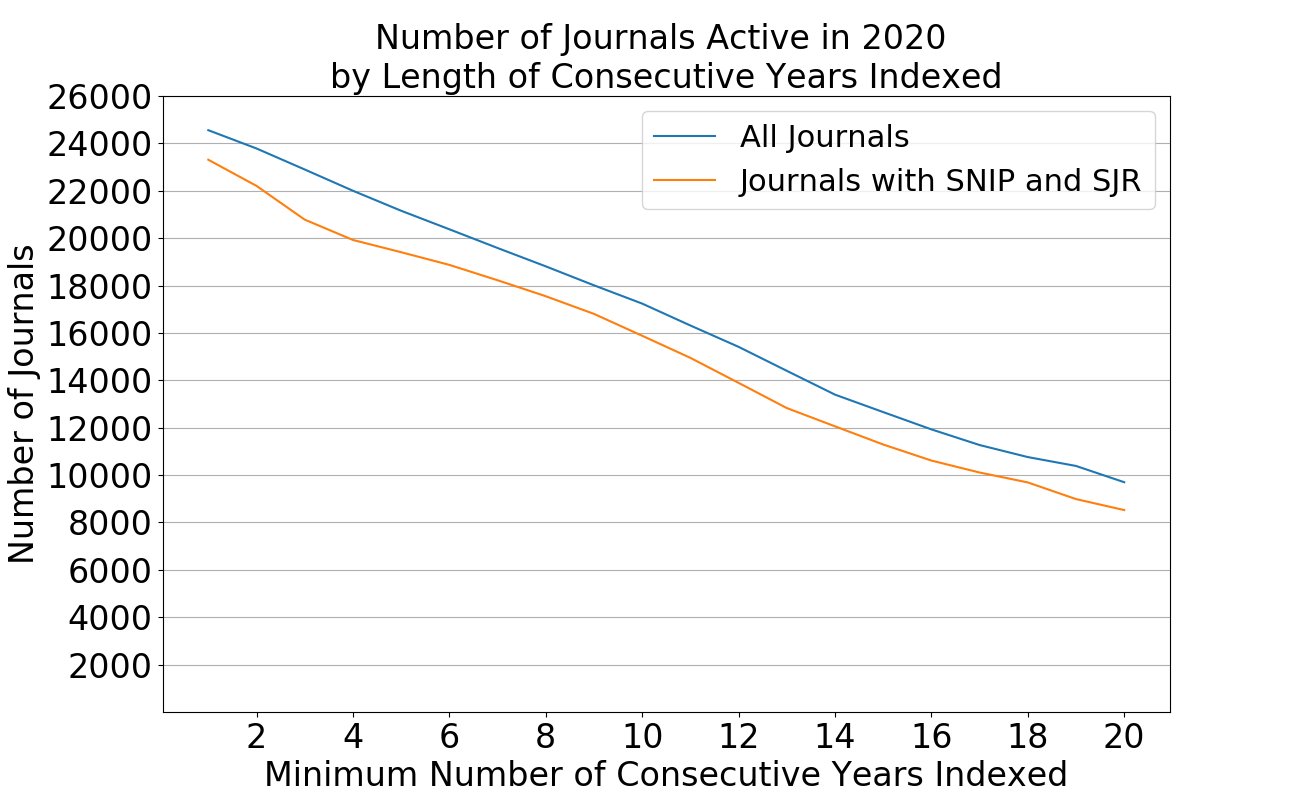}
	\caption{The number of journals active in 2020 is examined as a function of the amount of historical data (i.e., the number of indexed years) the journals possess. The plot shows how many journals in the dataset have at least the number of indexed years specified on the x-axis.} \label{fig:active_journals}
\end{figure}

The amount of historical data available for the journals in the dataset will play a role in the performance of the predictive models trained on the data. As many journals are relatively new or have been only recently indexed in Scopus, a large portion of the journals in the dataset will not have data for the full range of years from 2000 to 2020. In Figure \ref{fig:active_journals}, we show the number of journals in the dataset in terms of how many years of indexed data the journals have. The x-axis denotes a minimum requirement on the number of indexed years while the y-axis denotes the number of journals that satisfy this requirement. We also plot the same function with the additional requirement that the journals have SNIP and SJR values for their index years. As some journals are missing these values in some of their years, this number of journals is slightly lower. Since we use the SNIP and SJR values for our model training, the second plotted line is indicative of the actual amount of data that is available for our experiments.

\section{Predictive Models} \label{sec:models}

To set up our predictive models, we must make decisions on a number of items including input and output features, model configuration parameters and, where applicable, model architecture. In this section, we describe the feature selections that we consider and the primary configuration options of the predictive models. We cover both standard machine learning models, used primarily for comparative purposes, and the neural network models that we propose for use in practice.

\subsection{Output Feature Selection} \label{sec:output}

We begin with identification of the desired output features from our predictive models. Specification of these features is necessary in order to determine appropriate input features to feed into the models. In the case of predicting the number of citations, there is a single clear choice for the value to predict; namely, the number of citations received during the next calendar year across all publications. We denote the number of citations received during year $x$ as $c_x$.

For the task of predicting the CiteScore received during year $x$, denoted $cs_x$, there are multiple ways to approach the problem. The predictive model could, for instance, be designed to directly output $cs_x$. However, this fails to leverage past data that we already know with certainty. When the task is to predict $cs_x$ for the next calendar year, only the metrics for the final year of the CiteScore window (i.e., the number of publications and the number of citations received during this year) are unknown. Referring back to the CiteScore calculation of Formula \ref{eq:citescore}, all values except $c_{x,x}$, $c_{x-1,x}$, $c_{x-2,x}$, $c_{x-3,x}$ and $p_x$ will already be known from historical data. Thus, prediction of only the unknown values to plug into the calculation will result in a lower degree of error.

Prediction of the unknown citation values for the final year of the CiteScore window can be handled in two main ways. In the first, we predict the sum of the unknown values. This is the total number of citations received during year $x$ for publications released during the four year window, denoted $c_{x,w_4}$. The alternative is to predict each of the four values separately. For the later, we consider the use of a separate model for each of the four values as opposed to a single model responsible for predicting four output values simultaneously.

Finally, we must consider how to predict $p_x$, the number of publications in year $x$. Preliminary experiments have shown that a persistence prediction (i.e., repetition of the value from the previous year) performs on par for this task with the other predictive models (described in Subsections \ref{sec:standard_models} and \ref{sec:neural_net_models}) we have explored. Thus for the sake of computational efficiency, we use a persistence prediction by setting $p_x = p_{x-1}$.

\subsection{Input Feature Selection} \label{sec:input}

Using the datasets we have prepared, we have a variety of features available for use as input to the predictive models and a number of years over which these features can be drawn. The choice of which features to use should be tailored to the output value to be predicted. Furthermore, the number of years worth of historical data from which to draw these features must be determined. Care must be taken to not inundate the models with too many input features as this may lead to overfitting (i.e., poor ability of the model to generalize to new data due to having learned the training data distribution too closely).

\begin{figure}[h]
	\centering
	\includegraphics[width=0.7\textwidth]{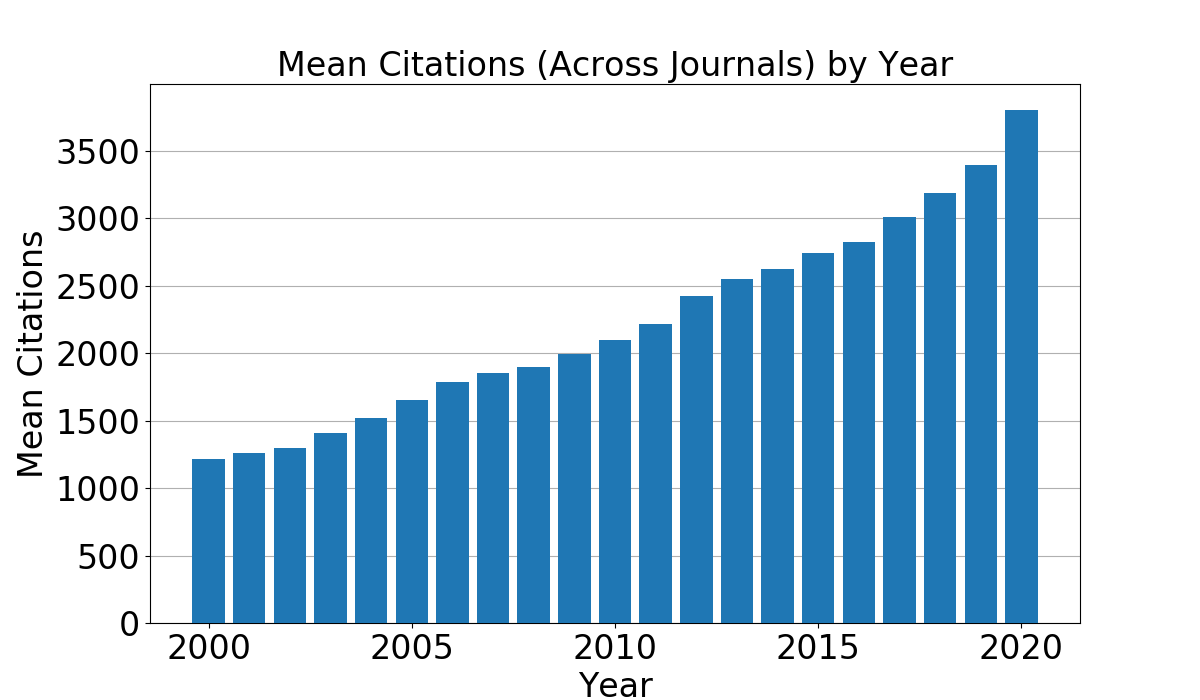}
	\caption{Mean citations received across all journals by each year covered in our dataset. This plot shows a monotonic increase in mean citations with each year.} \label{fig:citations_by_year}
\end{figure}

We first note that there is an increasing trend in average citations (see Figure \ref{fig:citations_by_year}) and publications over the years. Due to this, we provide the year for which the prediction is to be made as an input feature for all versions of the predictive tasks. Additionally, we consider the percent of publications not cited during the most recent year $x$ and the number of publications to always be a useful pieces of information and thus we use these features for all predictive tasks as well. For the remainder of the input features, we have made selections based on the output value to be predicted. For some predictive tasks, we consider multiple possible selections of input features. Through our model configuration experiments, described in Section \ref{sec:model_selection}, we identify the most promising input feature selections.

The notation used for all features is summarized in Table \ref{tab:notation}. The feature configurations (i.e., selections of input and output features) we have considered are detailed in a feature selection matrix shown in Table \ref{tab:features}. Each row of the matrix corresponds to a feature while each column corresponds to an overall selection of features for a particular predictive task. The bold horizontal rule separates the input features (above) from the output features (below). The first two selection columns correspond to two feature configurations for the task of predicting the citations received in the next year. The subsequent three columns correspond to configurations for the prediction of the sum of the citation values required for the CiteScore calculation. The remaining columns correspond to configurations for the prediction of each individual citation value required for the CiteScore calculation.

\begin{table} [!h] 
	\centering
	\begin{tabular}{ |l|l|l| }
		\hline
		\textbf{Notation} & \textbf{Description} & \textbf{Statistics} \\ \hline
		$x$ & The year $x$ & Range: 2000 - 2020  \\ \hline
		$p_x$ & \parbox[c]{5cm}{Number of publications\\during year $x$} & \parbox[c]{4cm}{Mean: 108\\Std dev: 282} \TBstrut\\ \hline
		$p_{x,w_y}$ & \parbox[c]{5cm}{Number of publications\\over $y$ years ending with\\year $x$ (inclusive)} & N/A \\ \hline
		$c_x$ & \parbox[c]{5cm}{Number of citations\\received during year $x$} & \parbox[c]{4cm}{Mean: 2880\\Std dev: 14,011} \TBstrut\\ \hline
		$c_{x,y}$ & \parbox[c]{5cm}{Number of citations\\received during year $x$ for\\publications from year $y$} & N/A \\ \hline
		$c_{x,w_y}$ & \parbox[c]{5cm}{Number of citations\\received during year $x$ for\\publications over the past $y$ years} & N/A \\ \hline
		$nc_x$ & \parbox[c]{5cm}{Percentage of publications\\from year $x$ not cited} & \parbox[c]{4cm}{Mean: 28.7146\\Std dev: 28.7169} \TBstrut\\ \hline
		$SNIP_x$ & SNIP value assigned for year $x$ & \parbox[c]{4cm}{Mean: 0.8099\\Std dev: 0.9634} \TBstrut\\ \hline 
		$SJR_x$ & SJR value assigned for year $x$ & \parbox[c]{4cm}{Mean: 0.6740\\Std dev: 1.2940} \TBstrut\\ \hline
	\end{tabular}
	\caption{Summary of the notation used for all input and output features. All features except $x$ are at the journal-level (i.e., feature values pertain to a specific journal). The third column provides basic statistics on the features measured across all journals and all years in the dataset. The high standard deviations are caused by distributions with long tails (see e.g., Figures \ref{fig:citations_distribution} and \ref{fig:publications_distribution}). Statistics on features involving two years as indices or a year and a window of time are omitted for simplicity.}
	\label{tab:notation}
\end{table}

\setlength\tabcolsep{4.5pt}
\begin{table} 
	\centering
	\begin{tabular}{ |c|c|c|c|c|c|c|c|c|c|c|c| }
		\hline
		& \rotatebox[origin=c]{90}{Citations Basic} & \rotatebox[origin=c]{90}{Citations Full} & \rotatebox[origin=c]{90}{CiteScore Sum Basic} & \rotatebox[origin=c]{90}{CiteScore Sum Detailed} & \rotatebox[origin=c]{90}{CiteScore Sum Full} & \rotatebox[origin=c]{90}{CiteScore Year 4} & \rotatebox[origin=c]{90}{CiteScore Year 3} & \rotatebox[origin=c]{90}{CiteScore Year 2 Basic} & \rotatebox[origin=c]{90}{CiteScore Year 2 Full}  & \rotatebox[origin=c]{90}{CiteScore Year 1 Basic}  & \rotatebox[origin=c]{90}{CiteScore Year 1 Full}  \\ \hline
		$x$ & X & X & X & X & X & X & X & X & X & X & X \\ \hline
		$nc_x$ & X & X & X & X & X & X & X & X & X & X & X \\ \hline
		$c_x$ & X & X & & & & & & & & &  \\ \hline
		$p_x$ & X & X & & X & X & X & X & & & &  \\ \hline
		$SNIP_x$ & & X & X & X & X & X & X & X & X & X & X \\ \hline
		$SJR_x$ & & X & X & X & X & X & X & X & X & X & X \\ \hline
		$c_{x,x}$ & & & & & X & X & X & & & &  \\ \hline
		$c_{x,x-1}$ & & & & & X & & X & X & X & &  \\ \hline
		$c_{x,x-2}$ & & & & & X & & & X & X & X & X \\ \hline
		$c_{x,x-3}$ & & & & & & & & & & X & X \\ \hline
		$c_{x-1,x-1}$ & & & & & X & & & & X & &  \\ \hline
		$c_{x-1,x-2}$ & & & & & X & & & & & & X \\ \hline
		$c_{x-2,x-2}$ & & & & X & X & & & & & & X \\ \hline
		$c_{x,w_4}$ & & & X & X & X & & & & & &  \\ \hline
		$c_{x,w_3}$ & & & X & X & & & & & & &  \\ \hline
		$c_{x-1,w_2}$ & & & & X & & & & & & &  \\ \hline
		$p_{x-1}$ & & & & X & X & & X & X & X & &  \\ \hline
		$p_{x-2}$ & & & & X & X & & & X & X & X & X \\ \hline
		$p_{x-3}$ & & & & & & & & & & X & X \\ \hline
		$p_{x,w_3}$ & & & X & & & & & & & & \\ \hline
		$p_{x,w_4}$ & & & X & X & X & & & & & & \\ \specialrule{.2em}{0em}{0em}
		$c_y$ & X & X & & & & & & & & & \\ \hline
		$c_{y,w_4}$ & & & X & X & X & & & & & &  \\ \hline
		$c_{y,y}$ & & & & & & X & & & & &  \\ \hline
		$c_{y,y-1}$ & & & & & & & X & & & &  \\ \hline
		$c_{y,y-2}$ & & & & & & & & X & X & &  \\ \hline
		$c_{y,y-3}$ & & & & & & & & & & X & X \\ \hline
	\end{tabular}
	\caption{Feature selection matrix for all input/output feature configurations tested. Rows correspond to features, with input features above the bold horizontal rule and output features below. Columns correspond to feature configurations.}
	\label{tab:features}
\end{table}

Beyond the selection of which input features are most appropriate for each predictive task, it is necessary to choose how many years worth of input features are to be provided. In the context of an LSTM, this determines the length of the sequences of timesteps given to the network as input. For all other models, the input is given as a single vector containing the input features from each year over the chosen number of input years. In other words, the total number of input features is the number of selected features times the number of years. When configuring our models, we consider selections of 3, 4, 5, 6, 8 or 10 years of historical data as input to predict for the next year.

\subsection{Standard Machine Learning Models} \label{sec:standard_models}

To establish the predictive abilities of standard machine learning algorithms, we apply four classic models: linear regression, decision tree, random forest and k-nearest neighbors.

Linear regression has no significant configuration parameters and thus only requires an appropriate selection of input and output features for its use in this context.

For the decision tree, we explore values of 3, 6, 9, 12 and 15 for the maximum allowable tree depth. The maximum tree depth places a limitation on how many levels the tree may have, thus limiting the complexity of the model. Higher values allow for higher capacity in the model, but also lead to larger model size, longer training time and an increased risk of overfitting. It is often practical to start with smaller values and assess whether a deeper tree is needed. We additionally explore values of 2, 5 and 10 for the number of minimum samples required to split an internal node. The selection of this value helps to control the quality of the splits at internal nodes. Low values may lead to decisions based on too few samples, leading to overfitting. Whereas high values may lead to the use of too many samples, limiting the predictive ability of the model.

As the random forest model is an ensemble of decision trees, it has similar parameters to configure. We explore the same ranges of values for the maximum allowable tree depth and minimum samples to split an internal node as with the decision tree model. In addition, we use values of 20, 50 and 100 for the number of trees to form the ensemble from. Similar to the tree depth, this parameter impacts the complexity of the model. Increasing the number of trees will often improve the quality of the results up to a certain point, at which there are diminishing returns. It is desirable to avoid using more trees than necessary as this has a large impact on the model size and training time.

For k-nearest neighbors, we explore choices of 10, 20, 50 and 100 for the value of k (i.e., the number of neighbors to use when performing interpolation). Setting this value too low can result in a model that is overly sensitive to a small set of neighbors, whereas setting the value too high can result in over-generalization by using too many neighbors. A balance must be found to achieve good performance from the model.

All combinations of model parameters and feature configurations are compared via grid search. Details and results of the comparisons are given in Section \ref{sec:model_selection}. 

We have additionally considered the use of a Poisson regression model. Since this model is only appropriate for regression of count data (i.e., natural numbers), it can only be applied to the citations prediction task and not the CiteScore prediction task. Experimentation with this model showed worse performance than the other models we have investigated as well as heuristic baselines (introduced in Section \ref{sec:experiments}). We therefore omit the results from this model in the interest of space.

Given that we are working with timeseries data, one might also consider the application of an ARIMAX model. However, ARIMAX is intended to learn from a single, generally lengthy, timeseries and predict its subsequent elements. In contrast, our setting involves training models from a large set of many small timeseries sequences (one per journal). A separate ARIMAX model could theoretically be trained for each journal, but with sequences that range from 3 to 10 (years) in length, this offers too little data for effective model training.

\subsection{Neural Network Models} \label{sec:neural_net_models}

We hypothesize that the best performance in the predictive tasks will be obtained through neural network models. When designing and training neural networks, a number of hyperparameters must be configured. With respect to the model architecture, it is necessary to specify the number of hidden layers to use, the size of each hidden layer and the activation functions to be used after each such layer. For the MLP, we consider architectures using 1, 2 or 4 hidden layers with all layers using a size of 50, 100 or 200 units and a ReLU activation function.

For the LSTM, we consider architectures using 1 or 2 hidden layers with all layers using a size of 25, 50 or 100 units and a tanh activation function. Since the LSTM maintains a hidden state which is updated as each timestep (i.e., year of historical data) is processed, it does not require as high of a capacity as the MLP which must process all years of input data simultaneously. For comparison, we additionally test a standard RNN across the same range of configuration settings as the LSTM.

The selection of the number of layers and the number of units per layer for any neural network impacts the overall capacity of the network. Each neuron in the network learns a set of weights that allows it to act as a detector of useful information in the input data. Increasing the number of units (i.e., neurons) in a layer provides the network with a greater capacity to detect useful information. Using multiple hidden layers builds a hierarchy of detections, allowing for a greater power of abstraction of information. Additionally, increasing the number of layers in a network naturally increases capacity due to the extra neurons introduced in the new layers. However, the use of too many layers or units per layer risks leading to overfitting. The identification of an appropriate architecture often requires experimentation to find configurations that perform well. As with the standard regression models, we perform a grid search (described in Section \ref{sec:model_selection}) over all combinations of feature configurations and model hyperparameter configurations to identify the best model configuration.

We train all of the models using the adam optimizer with a mean squared error objective function. We keep the default optimizer hyperparameters from the libraries we have used (scikit-learn\footnote{\url{https://scikit-learn.org/stable/}} for the MLP and Tensorflow\footnote{\url{https://www.tensorflow.org/}} for the LSTM and standard RNN). We allow the models to train for an unbounded number of epochs until an early stopping process determines that no further improvements are occurring. For the MLP, this process is part of the default implementation and we do not modify its parameterization. The LSTM and RNN do not support this by default, thus we add an early stopping callback that triggers after 15 epochs of no further improvements over the best epoch validation score recorded during training.

\section{Model Configuration Experiments} \label{sec:model_selection}

To select the final configuration to use for each model from the options laid out in Section \ref{sec:models}, we compare the performance of each configuration using a grid search. In this section, we describe the data preparation used for these experiments, the comparison criteria we apply, and the best performing configurations that we have identified.

\subsection{Data Preparation} \label{sec:data_preparation}

%Due to a large variation in typical yearly citation counts across different journals (ranging from values in the dozens up to values in the tens of thousands), preliminary experiments have revealed that predictions for journals with low citation counts suffer from a disproportionate degree of error. Due to this, we partition the dataset and train two different versions of each model - one for lower citation journals and one for higher citation journals. The low citation dataset is formed by ranking each journal by its highest yearly citation value across all years it has been indexed and removing the top 30\% of the journals. The high citation version is similarly formed by removing the bottom 30\% based on their lowest yearly citation value.

We first split the data into training and testing partitions for the models by selecting 90\% of the journals at random for the training partition and using the remaining 10\% for the testing partition. Once partitioned, we produce the input features for the models by passing a sliding window over the yearly data for each journal (ordered chronologically) where the window is sized according to the number of years of historical data to be used for the input features. Each positioning of the window must leave space for an additional year of data after the window to serve as the ground truth for the output value to be predicted. Thus, for a window of $x$ years of input data, $x+1$ total years worth of data is required to form one data sample. We slide the window using a stride of 1 in order to maximize the number of samples produced from the dataset.

\begin{figure} 
	\centering
	\includegraphics[width=0.7\textwidth]{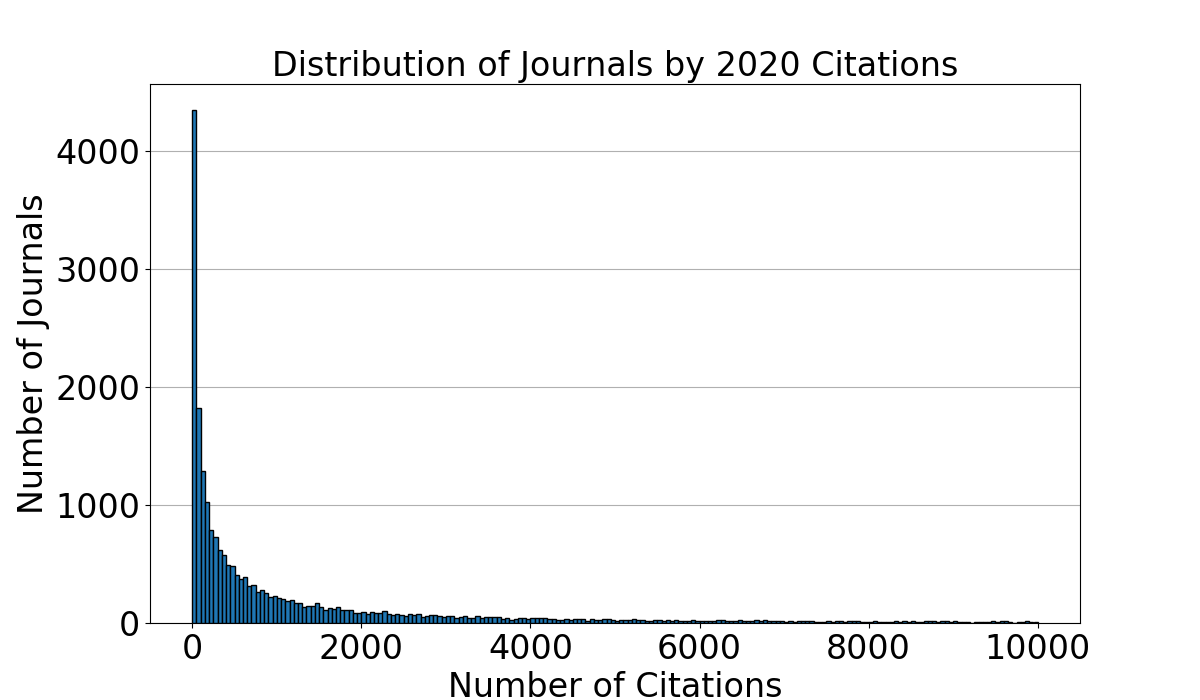}
	\caption{Distribution of journals by the number of citations received during 2020. For ease of visualization, the full tail of the distribution is not shown. The maximum number of citations received in 2020 across the dataset journals is 832,260.} \label{fig:citations_distribution}
\end{figure}

\begin{figure} 
	\centering
	\includegraphics[width=0.7\textwidth]{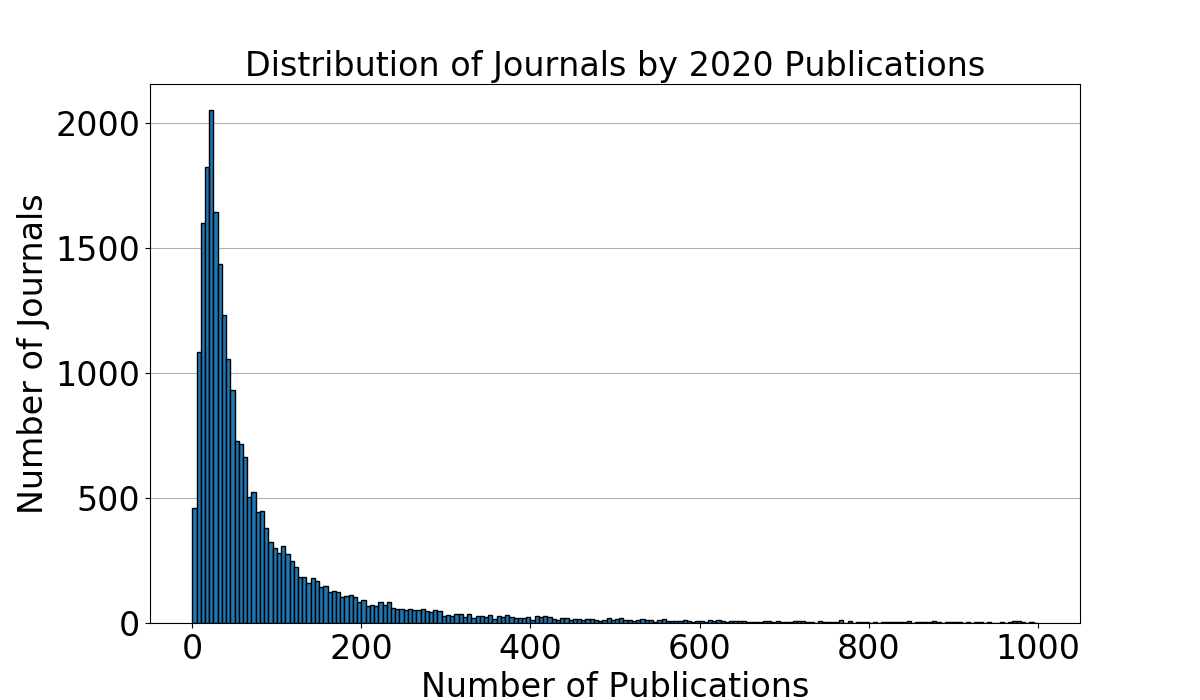}
	\caption{Distribution of journals by the number of publications made during 2020.  For ease of visualization, the full tail of the distribution is not shown. The maximum number of publications released in 2020 across the dataset journals is 21,190.} \label{fig:publications_distribution}
\end{figure}

The majority of the input features have distributions that are heavily skewed towards smaller values. Examples of this can be seen in the distribution of citations received by journals in 2020 (Figure \ref{fig:citations_distribution}) and the number of publications made by journals in 2020 (Figure \ref{fig:publications_distribution}). For all input features with this property, we preprocess the data by applying a power transform in order to give the data the similarity of a normal distribution. This is done for all input features except for the year and percent not cited features, which we instead normalize to a range of $[0,1]$.

\subsection{Configuration Selection} \label{sec:config_selection}

To identify the best configuration of a model from the possible choices we have laid out, we perform a grid search and select the best performing variant. To do so, we must define the criterion through which to assess performance. We are primarily interested in minimization of prediction errors. However, as observable in the distribution shown in Figure \ref{fig:citations_distribution}, there is significant disparity between yearly citation values across different journals. Due to this, a margin of error of 100 citations has very different implications depending on the ground truth for the sample (e.g., 50\% of 200 citations versus 5\% of 2000 citations). We therefore consider it to be more meaningful to interpret error proportional to the ground truth values. To this end, we measure the performance of the models with respect to mean absolute percent error, where lower values are indicative of better performance.

For each model, we execute the grid search over the feature selections and sequence lengths laid out in Subsections \ref{sec:output} and \ref{sec:input} and the model configuration parameters laid out in Subsections \ref{sec:standard_models} and \ref{sec:neural_net_models}. For each model configuration, we perform 10-fold cross-validation \cite{39} on the training partition of the data. This is done by splitting the data into 10 partitions of roughly equal size. Across 10 iterations, each one of the partitions is set aside to use for testing while the other 9 are used for training the model under the selected configuration. We measure performance as the average of the test results across the 10 iterations. The use of cross-validation provides a robust measure of performance by ensuring the model is exposed to a variety of testing data. Furthermore, the averaging of results across multiple iterations mitigates the impact of stochastic aspects of training algorithms on the results of the experiments.

\begin{table*} 
	\small
	\centering
	\begin{tabular}{ | c | c | c | c | c | c | c | }
		\hline
		\parbox[t]{2.1cm}{\textbf{Linear}\\\textbf{Regression}} & \parbox[t]{1.9cm}{\textbf{Decision}\\\textbf{Tree}} & \parbox[t]{1.8cm}{\textbf{Random}\\\textbf{Forest}} & \parbox[t]{2.0cm}{\textbf{k-Nearest}\\\textbf{Neighbors}} & \textbf{MLP} & \textbf{RNN} & \textbf{LSTM} \\ \hline
		12 & 180 & 540 & 48 & 108 & 72 & 72  \\ \hline
	\end{tabular}
	\caption{Number of configurations tested for the citations prediction task.}
	\label{tab:config_counts_citations}
	\vspace{0.8cm}
\end{table*}

\begin{table*} 
	\small
	\centering
	\begin{tabular}{ | c | c | c | c | }
		\hline
		\parbox[t]{2.6cm}{\textbf{Linear}\\\textbf{Regression}} & \textbf{Decision Tree} & \textbf{Random Forest} & \parbox[t]{2.6cm}{\textbf{k-Nearest}\\\textbf{Neighbors}} \\ \hline
		\parbox[t]{2.6cm}{Years: 10\\Features: Basic} & \parbox[t]{2.8cm}{Years: 10\\Features: Basic\\Max depth: 9\\Min samples: 10} & \parbox[t]{2.8cm}{Years: 10\\Features: Basic\\Max depth: 15\\Min samples: 5\\Num trees: 100} & \parbox[t]{2.6cm}{Years: 3\\Features: Basic\\k: 20} \\ \hline
	\end{tabular}
	\caption{Results of the model configuration experiments for the citations prediction task using standard machine learning models. Each column shows the best-performing configuration of a model.}
	\label{tab:config_results_citations_classic}
	\vspace{0.8cm}
\end{table*}

\begin{table*} 
	\small
	\centering
	\begin{tabular}{ | c | c | c | }
		\hline
		\textbf{MLP} & \textbf{RNN} & \textbf{LSTM} \\ \hline
		\parbox[t]{2.6cm}{Years: 5\\Features: Full\\\# Layers: 2\\Layer Size: 200} & \parbox[t]{2.6cm}{Years: 8\\Features: Full\\\# Layers: 1\\Layer Size: 50} & \parbox[t]{2.6cm}{Years: 10\\Features: Full\\\# Layers: 1\\Layer Size: 25} \\ \hline
	\end{tabular}
	\caption{Results of the model configuration experiments for the citations prediction task using neural network models. Each column shows the best-performing configuration of a model.}
	\label{tab:config_results_citations_neural_net}
	\vspace{0.8cm}
\end{table*}

\begin{table*} 
	\small
	\centering
	\begin{tabular}{ | c | c | c | c | c | c | c | }
		\hline
		\parbox[t]{2.1cm}{\textbf{Linear}\\\textbf{Regression}} & \parbox[t]{1.9cm}{\textbf{Decision}\\\textbf{Tree}} & \parbox[t]{1.8cm}{\textbf{Random}\\\textbf{Forest}} & \parbox[t]{2.0cm}{\textbf{k-Nearest}\\\textbf{Neighbors}} & \textbf{MLP} & \textbf{RNN} & \textbf{LSTM} \\ \hline
		54 & 810 & 2,430 & 216 & 486 & 324 & 324  \\ \hline
	\end{tabular}
	\caption{Number of configurations tested for the CiteScore prediction task.}
	\label{tab:config_counts_cs}
	\vspace{0.8cm}
\end{table*}

\begin{table*}
	\small
	\centering
	\begin{tabular}{ | c | c | c | c | c | }
		\hline
		& \parbox[t]{2.6cm}{\textbf{Linear}\\\textbf{Regression}} & \textbf{Decision Tree} & \textbf{Random Forest} & \parbox[t]{2.6cm}{\textbf{k-Nearest}\\\textbf{Neighbors}} \\ \hline
		
		\textbf{Sum} & \parbox[t]{2.6cm}{Years: 10\\Features: Full} & \parbox[t]{2.8cm}{Years: 8\\Features: \\Detailed\\Max depth: 9\\Min samples: 10} & \parbox[t]{2.8cm}{Years: 10\\Features: \\Detailed\\Max depth: 15\\Min samples: 10\\Num trees: 100} & \parbox[t]{2.6cm}{Years: 3\\Features: Full\\k: 20} \\ \hline
		
		\textbf{Year 1} & \parbox[t]{2.6cm}{Years: 10\\Features: Full} & \parbox[t]{2.8cm}{Years: 8\\Features: Full\\Max depth: 9\\Min samples: 2} & \parbox[t]{2.8cm}{Years: 10\\Features: Full\\Max depth: 15\\Min samples: 10\\Num trees: 100} & \parbox[t]{2.6cm}{Years: 3\\Features: Full\\k: 20} \\ \hline
		
		\textbf{Year 2} & \parbox[t]{2.6cm}{Years: 10\\Features: Full} & \parbox[t]{2.8cm}{Years: 8\\Features:\\ Basic\\Max depth: 9\\Min samples: 2} & \parbox[t]{2.8cm}{Years: 10\\Features: Full\\Max depth: 15\\Min samples: 2\\Num trees: 100} & \parbox[t]{2.6cm}{Years: 3\\Features: Basic\\k: 20} \\ \hline
		
		\textbf{Year 3} & \parbox[t]{2.6cm}{Years: 10\\Features: Basic} & \parbox[t]{2.8cm}{Years: 8\\Features: \\Basic\\Max depth: 9\\Min samples: 10} & \parbox[t]{2.8cm}{Years: 8\\Features: Basic\\Max depth: 15\\Min samples: 5\\Num trees: 100} & \parbox[t]{2.6cm}{Years: 3\\Features: Basic\\k: 20} \\ \hline
		
		\textbf{Year 4} & \parbox[t]{2.6cm}{Years: 10\\Features: Basic} & \parbox[t]{2.8cm}{Years: 3\\Features: \\Basic\\Max depth: 9\\Min samples: 10} & \parbox[t]{2.8cm}{Years: 3\\Features: Basic\\Max depth: 12\\Min samples: 2\\Num trees: 100} & \parbox[t]{2.6cm}{Years: 3\\Features: Basic\\k: 100} \\ \hline
	\end{tabular}
	\caption{Results of the model configuration experiments for the CiteScore prediction task using standard machine learning models. The first row corresponds to the task of predicting the sum of the citation values needed to calculate the CiteScore (columns 3-5 of Table \ref{tab:features}) while the other rows correspond to the tasks of predicting the citations from publications of a specific year in the CiteScore window (columns 6-11 of Table \ref{tab:features}). Each column shows the best-performing configuration of a model.}
	\label{tab:config_results_cs_classic}
	\vspace{0.8cm}
\end{table*}

\begin{table*}
	\small
	\centering
	\begin{tabular}{ | c | c | c | c | }
		\hline
		 & \textbf{MLP} & \textbf{RNN} & \textbf{LSTM} \\ \hline
		
		\textbf{Sum} & \parbox[t]{3.1cm}{Years: 6\\Features: Detailed\\\# Layers: 2\\Layer Size: 50} & \parbox[t]{2.6cm}{Years: 10\\Features: Full\\\# Layers: 1\\Layer Size: 25} & \parbox[t]{2.6cm}{Years: 6\\Features: Full\\\# Layers: 1\\Layer Size: 50} \\ \hline
		
		\textbf{Year 1} & \parbox[t]{3.1cm}{Years: 8\\Features: Full\\\# Layers: 1\\Layer Size: 50} & \parbox[t]{2.6cm}{Years: 6\\Features: Full\\\# Layers: 1\\Layer Size: 100} & \parbox[t]{2.6cm}{Years: 10\\Features: Full\\\# Layers: 1\\Layer Size: 25} \\ \hline
		
		\textbf{Year 2} & \parbox[t]{3.1cm}{Years: 8\\Features: Basic\\\# Layers: 4\\Layer Size: 50} & \parbox[t]{2.6cm}{Years: 10\\Features: Full\\\# Layers: 1\\Layer Size: 50} & \parbox[t]{2.6cm}{Years: 10\\Features: Full\\\# Layers: 2\\Layer Size: 50} \\ \hline
		
		\textbf{Year 3} & \parbox[t]{3.1cm}{Years: 6\\Features: Basic\\\# Layers: 4\\Layer Size: 50} & \parbox[t]{2.6cm}{Years: 10\\Features: Basic\\\# Layers: 1\\Layer Size: 25} & \parbox[t]{2.6cm}{Years: 10\\Features: Basic\\\# Layers: 2\\Layer Size: 25} \\ \hline
		
		\textbf{Year 4} & \parbox[t]{3.1cm}{Years: 4\\Features: Basic\\\# Layers: 1\\Layer Size: 50} & \parbox[t]{2.6cm}{Years: 5\\Features: Basic\\\# Layers: 1\\Layer Size: 100} & \parbox[t]{2.6cm}{Years: 6\\Features: Basic\\\# Layers: 2\\Layer Size: 25} \\ \hline
	\end{tabular}
	\caption{Results of the model configuration experiments for the CiteScore prediction task using neural network models. The first row corresponds to the task of predicting the sum of the citation values needed to calculate the CiteScore (columns 3-5 of Table \ref{tab:features}) while the other rows correspond to the tasks of predicting the citations from publications of a specific year in the CiteScore window (columns 6-11 of Table \ref{tab:features}). Each column shows the best-performing configuration of a model.}
	\label{tab:config_results_cs_neural_net}
	\vspace{0.8cm}
\end{table*}

The numbers of configurations tested during the grid search are shown in Table \ref{tab:config_counts_citations} for the citations prediction task (i.e., predicting the number of citations received during the next year) and the best model configurations identified are shown in Tables \ref{tab:config_results_citations_classic} and \ref{tab:config_results_citations_neural_net}. The configurations of input and output features included in the grid search correspond to the first two columns of Table \ref{tab:features}. The same information for the CiteScore prediction task (i.e., predicting the CiteScore received during the next year) is provided in Tables \ref{tab:config_counts_cs} - \ref{tab:config_results_cs_neural_net}. The configurations of input and output features used in the grid search for this task correspond to the nine CiteScore columns of Table \ref{tab:features}. A direct comparison between each of the selected configurations is provided in Section \ref{sec:experiments}.

Inspection of the parameters used by the selected model configurations helps to assess the appropriateness of the range of values tested. For the decision tree, the selection of a maximum tree depth of 9 in all selected configurations suggests that this value provides consistency in the performance of the model. That the value is situated in the middle of the tested range suggests that the model did not suffer from options that were too low or too high. On the other hand, the values selected for the minimum number of samples required to split internal nodes are drawn from both ends of the tested range. This suggests that this parameter had little practical impact in this application of the model as long as the values were drawn from a reasonable range.
	
The random forest configurations used the highest values available for the number of trees in all cases except one and used the highest available tree depth in all cases. This indicates that the models were able to fully make use of the available capacity under these high settings and that there is potential for further improvement by raising them higher. We have performed additional exploration (results omitted for brevity) to verify that while the values can indeed be raised higher, the change in performance is negligible while the increases in training time and model size are notable. As with the decision tree, the selection of the value for the minimum number of samples required to split internal nodes appears to have little impact on the model performance.

The k-nearest neighbors model consistently selected 20 as the best value for k with the exception of the configuration for the Year 4 CiteScore predictions, where a value of 100 was selected. Thus in general, a low value for the number of neighbors was more beneficial, although the model never selected the lowest available value of 10, suggesting that to drop the value much more would lead to a neighborhood lacking sufficient information to interpolate from. The use of a value of 100 in the Year 4 CiteScore model is likely due to the lower degree of information available to this model. Unlike the models for publications from other years of the CiteScore window, the input data for the Year 4 model contains no historical information pertaining to the publications for this year since those publications have not yet occurred. In the absence of this information, the model likely identifies better performance in the use of a large neighborhood to interpolate within, moving more towards highly generalized predictions to obtain better average case performance.

The neural networks settled on a number of different variations of architecture. This demonstrates that different possible architectures can be effectively used, provided that the network is given an appropriate capacity. Since none of the configurations fully made use of the capacity available among the choices of architecture, this suggests that the networks did not suffer from too little capacity. Although some configurations used the lowest available capacity, the selection of other network configurations trained for similar tasks using a higher capacity suggests that using such a low capacity is not necessary for the predictive tasks at hand.

While further tuning of model parameters by using smaller intervals between the tested settings may lead to slight improvements in performance, too fine of a granularity may also lead to overfitting of the parameters. One may additionally explore the sensitivity of model performance to the selection of the parameter values to further assess the appropriateness of their selection. This type of fine-tuning falls outside the scope of our work; we rely on the k-fold cross validation methodology to ensure the robustness of our results.

\section{Experimental Comparisons} \label{sec:experiments}

With the configuration of the predictive models finalized, we turn to the assessment of their practical applicability. To verify that the models offer useful predictions, we define a set of heuristic baseline predictions. We first compare the overall performance of the models and baselines. To provide a finer grained examination of the performance of the models, we then bucketize the test data and graph the performance of the models across the buckets. We provide analysis and discussion of the results we have observed.

For all experiments in this section, we follow the same steps of data preparation as described in Subsection \ref{sec:data_preparation}. To avoid any contamination of testing data with records used for the configuration of the model, we do not perform cross-validation at this stage of the experiments. Instead, we train the models on the full training partition and measure performance on the testing partition. We still take the overall performance as the average across 10 runs in order to help mitigate any impacts of stochasticity. Furthermore, since the LSTM uses information from the testing data for early stopping, we do not apply the early stopping procedure during these experiments for fairness of comparison. Instead, we train the model for a fixed number of epochs calculated as the average number of epochs used by the selected configuration from Subsection \ref{sec:config_selection}.

\subsection{Prediction Baselines}

We define three heuristic baselines to compare against the predictive models. The formulae we provide below are written in the context of predicting the total citations received in the next year. For the other predictive tasks we consider, they are defined analogously.

\medskip
\noindent \textbf{Persistence Baseline:} The first baseline is a persistence prediction in which the value to be predicted is assumed to be the same as its value from the last known element of the series. This is most appropriate for instances where there are relatively small changes expected between consecutive elements of the series. Given that journal performance often changes slowly from year to year, this is a reasonable choice. In our setting, this implies repetition of the final known year of data, defined as follows:

\begin{equation} \label{eq:persistence}
	c_y = c_{y-1}.
\end{equation}

\noindent \textbf{Delta Baseline:} The second baseline calculates the difference between the last and second last elements of the series and adds this difference to the last element to extrapolate the next value. This is most appropriate for instances where the series follows a relatively linear trend. This can capture stable trends of improvement, decline or stagnation in journal performance that persist over multiple years. The calculation is given as follows:

\begin{equation}
	c_y = c_{y-1} + \frac{c_{y-1} + c_{y-2}}{2}.
\end{equation}

\noindent \textbf{Weighted Delta Baseline:} The final heuristic baseline extends the concept of applying the difference between the final known elements by using a weighted sum of differences over the last five elements of the series. Journal performance may show minor fluctuations from year to year while following an overall trend better observable across a span of multiple years. This variant has the potential to better cope with these fluctuations while capturing the overall trend. We have selected weights that monotonically decrease from the most recent year to the least recent year in order to place greater importance on the recent behaviour of the journal, which is expected to be more relevant than behvaiour from older years. The heuristic is defined as follows:

\begin{equation}
	\begin{split}
		c_y = c_{y-1} + 0.4 * (c_{y-1} + c_{y-2}) + 0.3 * (c_{y-2} + c_{y-3}) + \\
		0.2 * (c_{y-3} + c_{y-4}) + 0.1 * (c_{y-4} + c_{y-5}).
	\end{split}
\end{equation}

\subsection{Model Performance Summary Statistics}

We begin by comparing all finalized models and baselines on their performance with respect to the mean absolute error (MAE), median absolute error (MedAE), mean absolute percent error (MAPE) and median absolute percent error (MedAPE) of the predictions made. In all four measures of error, lower values are indicative of better performance. We additionally compare the models in terms of their coefficients of determination ($R^2$). The coefficient of determination of a regression model is a measure of the proportion of the variance in values to be predicted that is explained by the model. This measure is reflective of the improvement gained through the use of the model over using the mean value of the data for every prediction. For a vector $Y$ of $n$ ground truth values to be predicted and a vector $\hat{Y}$ of $n$ corresponding predictions, the coefficient of determination is defined as:

\begin{equation}
	R^2 = 1 - \frac{\sum\limits_{i=1}^n (Y_i-\hat{Y}_i)^2}{\sum\limits_{i=1}^n (Y_i-\bar{Y})^2},
\end{equation}

where $\bar{Y}$ is the mean of the values in $Y$. The measure is upper-bounded by a value of 1, reflecting a perfect explanation of all variance by the model. A value of 0 indicates no improvement compared to the use of the mean value while negative values indicate worse performance than through the use of the mean value.

Results are given in Table \ref{tab:citation_performance} for citation predictions and Table \ref{tab:citescore_performance} for CiteScore predictions. For both predictive tasks, the LSTM achieves the best performance across all measures, with the exception of the $R^2$ measure for the citations prediction where the weighted delta baseline provides the best result with the LSTM providing the second-best result. The standard RNN offers performance very close to that of the LSTM with measures of error that are marginally worse. This is expected, given the recurrent nature of both models. For longer sequences of input, an LSTM would likely provide a greater gain in performance over a standard RNN. The positive performance of the two recurrent neural network models demonstrates the strong predictive abilities achieved through the use of non-linearity in the models augmented by the sequential processing of a timeseries of journal data while preserving an internal state on past sequence elements.

\begin{table} [!htp]
	\small
	\centering
	\begin{tabular}{ | c | c | c | c | c | c | }
		\hline
		\textbf{Predictive Model} & \textbf{MAE} & \textbf{MedAE} & \textbf{MAPE} & \textbf{MedAPE} & $\bm{R^2}$ \\ \hline
		Persistence Baseline & 426.141 & 105.000 & 12.53 & 9.38 & 0.9935 \\ \hline
		Delta Baseline & 349.180 & 106.609 & 16.43 & 9.18 & 0.9973 \\ \hline
		Weighted Delta Baseline & 281.875 & 82.592 & 12.18 & 7.16 & \textbf{0.9981} \\ \hline
		Linear Regression & 286.117 & 73.161 & 10.24 & 6.41 & 0.9964 \\ \hline
		Decision Tree & 351.301 & 77.679 & 11.06 & 6.98 & 0.9869 \\ \hline
		Random Forest & 249.680 & 68.696 & 10.32 & 6.29 & 0.9967 \\ \hline
		k-Nearest Neighbors & 490.424 & 95.465 & 13.45 & 8.88 & 0.9862 \\ \hline
		MLP & 292.378 & 71.828 & 10.33 & 6.40 & 0.9968 \\ \hline
		RNN & 267.999 & 67.205 & 10.07 & 6.18 & 0.9953 \\ \hline
		LSTM & \textbf{246.787} & \textbf{67.180} & \textbf{9.51} & \textbf{5.84} & 0.9975 \\ \hline
	\end{tabular}
	\caption{Comparison of performance statistics across the studied models for the citations prediction task. With the four measures of error, lower values indicate better performance. For the measure of $R^2$, higher values indicate better performance.}
	\label{tab:citation_performance}
	\vspace{0.8cm}
\end{table}

\begin{table} [!htp]
	\small
	\centering
	\begin{tabular}{ | c | c | c | c | c | c | }
		\hline
		\textbf{Predictive Model} & \textbf{MAE} & \textbf{MedAE} & \textbf{MAPE} & \textbf{MedAPE} & $\bm{R^2}$ \\ \hline
		Persistence Baseline & 0.279 & 0.153 & 11.07 & 7.47 & 0.9865 \\ \hline
		Delta Baseline & 0.330 & 0.190 & 15.30 & 9.03 & 0.9832 \\ \hline
		Weighted Delta Baseline & 0.293 & 0.162 & 12.28 & 7.81 & 0.9855 \\ \hline
		Linear Regression & 0.236 & 0.129 & 9.48 & 6.23 & 0.9902 \\ \hline
		Decision Tree & 0.255 & 0.142 & 10.30 & 6.86 & 0.9883 \\ \hline
		Random Forest & 0.235 & 0.132 & 9.62 & 6.26 & 0.9901 \\ \hline
		k-Nearest Neighbors & 0.312 & 0.160 & 11.41 & 7.85 & 0.9790 \\ \hline
		MLP & 0.219 & \textbf{0.123} & 9.13 & 5.89 & 0.9914 \\ \hline
		RNN &  0.219 & 0.124 & 9.14 & 5.94 & 0.9914 \\ \hline
		LSTM & \textbf{0.215} & \textbf{0.123} & \textbf{9.04} & \textbf{5.79} & \textbf{0.9919} \\ \hline
	\end{tabular}
	\caption{Comparison of performance statistics across the studied models for the CiteScore prediction task. With the four measures of error, lower values indicate better performance. For the measure of $R^2$, higher values indicate better performance.}
	\label{tab:citescore_performance}
	\vspace{0.8cm}
\end{table}

The MLP, random forest and linear regression models also provide reasonably good levels of performance. The random forest achieves a slight edge over the other two models for the citations prediction task while the MLP performs better on the CiteScore predictions task. The decision tree, while notably poorer in performance than the aforementioned models, achieves some improvement over the baseline methods. Finally, the k-nearest neighbors model consistently performs the worst among the models and in some instances performs worse than the baselines.

Between the three baseline methods tested, the weighted delta baseline achieves the best performance for prediction of citations while the persistence baseline performs best for CiteScore predictions. Interestingly, the weighted delta baseline achieves a lower mean error than both the MLP and linear regression for the citations prediction task, despite performing worse in the other measures of error. This is due to the heavy influence of the absolute errors on samples with larger magnitude citation values. This highlights the importance of evaluating the models under a variety of performance measures.

The $R^2$ values for all models appear unusually high, particularly in the case of the baselines where one might expect much worse performance from such simple heuristic approaches to prediction. This is due to the distributions of values to be predicted, which cover a wide range of values but are heavily skewed towards smaller values. The long tails of the distributions into large values cause the mean values of the sampled data to greatly overestimate the values at the densest portion of the distribution (refer back to Figure \ref{fig:citations_distribution} for an example). Due to this, every model appears to perform very well in its $R^2$ value as significant improvements over use of the mean as the predicted value are easily achieved. Thus, while the $R^2$ values can still be used to compare the relative performance of the different models, we conclude that the high scores should not be interpreted as indicating incredibly good performance from every model.

It is also interesting to note that the median errors for all models are consistently lower than their mean errors by a significant margin. This suggests a skewed distribution of errors in which a relatively small proportion of predicted values have larger errors than the rest, leading to an inflated mean value compared to the median. We therefore turn next to a closer inspection of model performance across bucketizations of the testing samples in order to better understand the predictive abilities of the models.

\subsection{Detailed View of Model Performance}

Given the wide range of citation and CiteScore values across the journals compounded by heavily skewed distributions, it is important to verify how the models perform across the distribution in addition to assessment based on the summary statistics presented thus far. To this end, we break the testing dataset up into buckets based on the ground truth values of the samples and measure the mean absolute error of predictions within each bucket. We compare the machine learning models as well as the best baseline for both predictive tasks. In the interest of space in the plotted data, we omit the results for the standard RNN as one would be expected to always use the LSTM instead.

\begin{figure*} [!htp]
	\centering
	\includegraphics[width=1\textwidth]{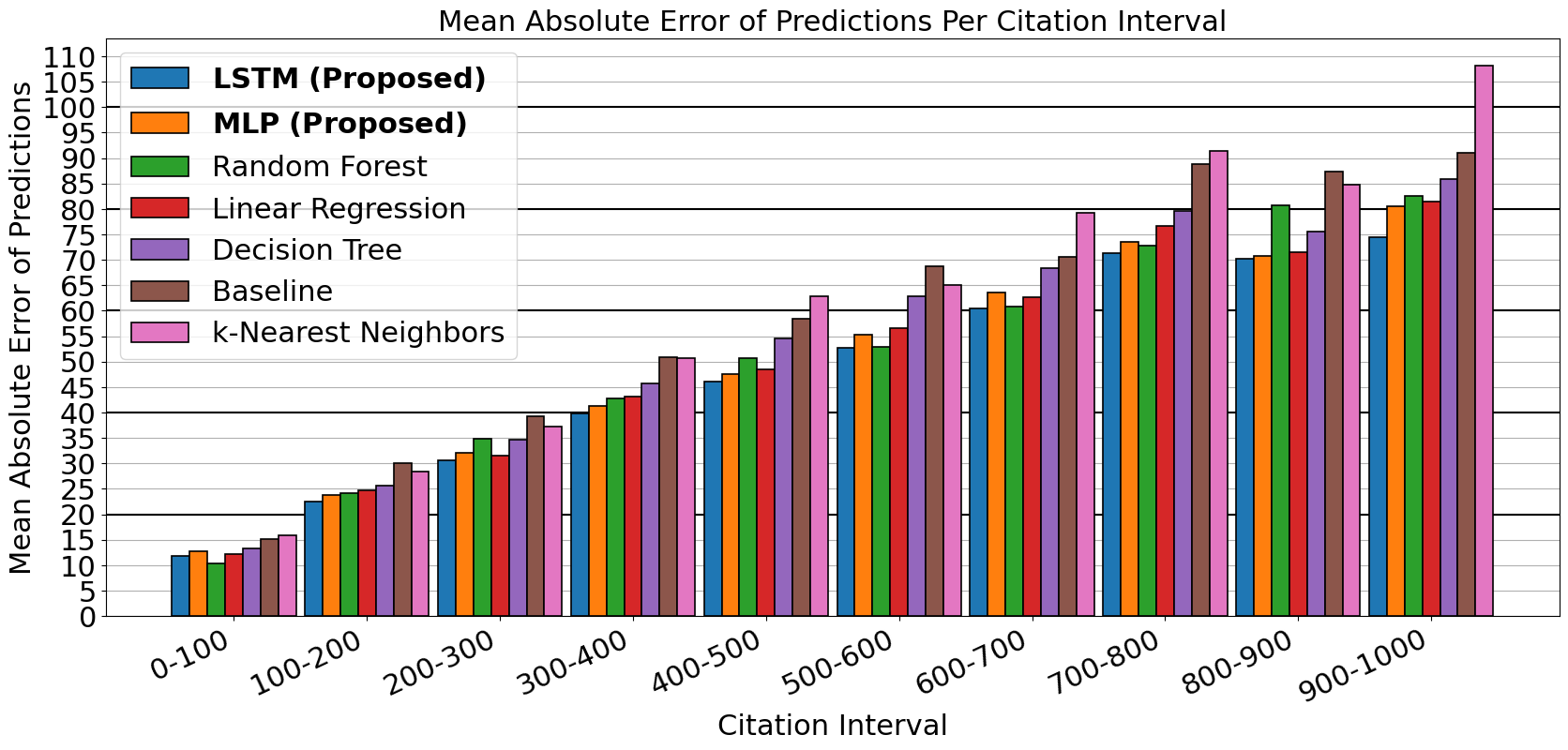}
	\caption{Mean absolute error of the selected model configurations for the citations prediction task. Test samples with a ground truth citation value in the range of 0-1,000 have been grouped into 10 buckets.} \label{fig:cite_small}
\end{figure*}

\begin{figure*} [!htp]
	\centering
	\includegraphics[width=1\textwidth]{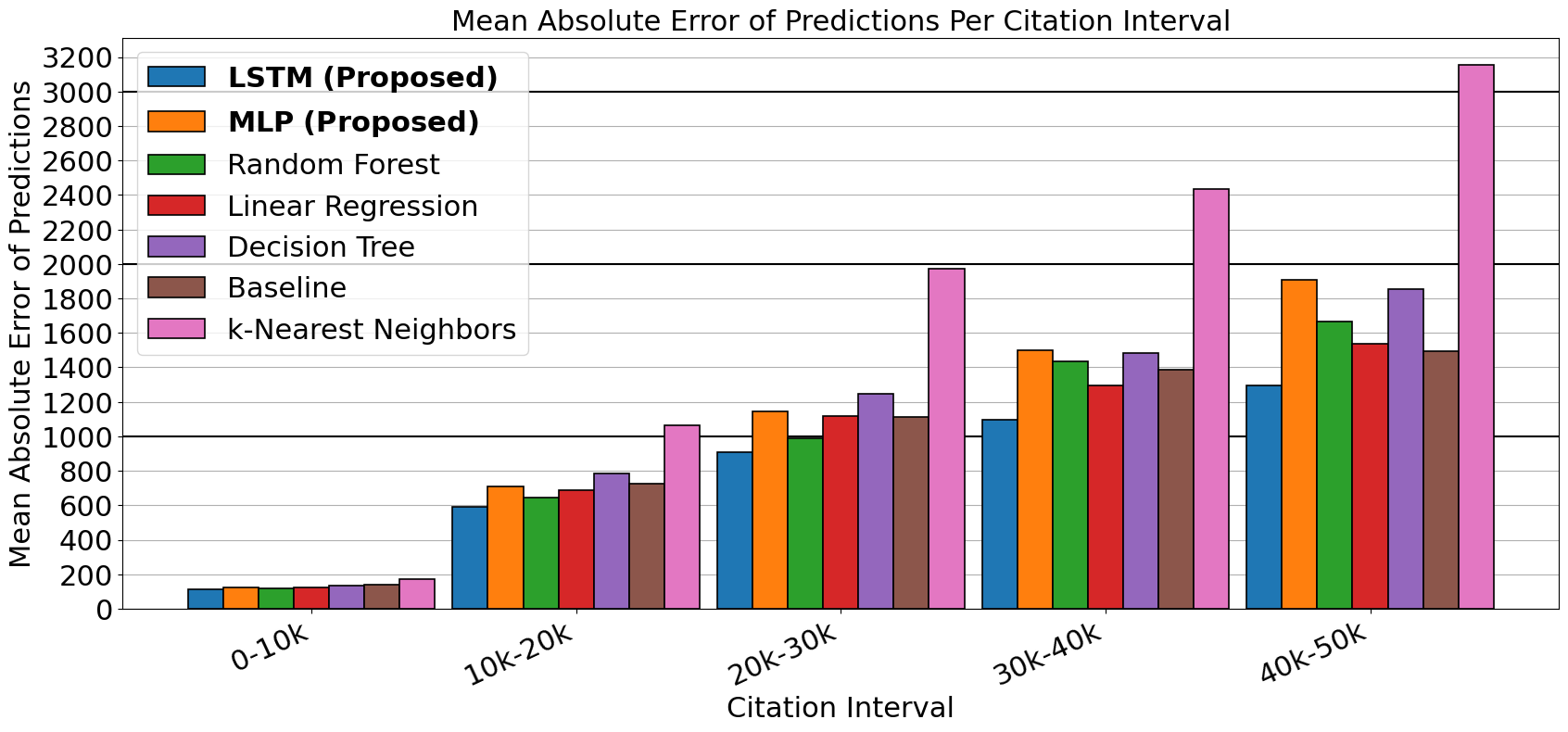}
	\caption{Mean absolute error of the selected model configurations for the citations prediction task. Test samples with a ground truth citation value in the range of 0-50,000 have been grouped into 5 buckets.} \label{fig:cite_large}
\end{figure*}

For the citations prediction task, we first focus on the range of 0-1,000 citations (Figure \ref{fig:cite_small}) broken into 10 buckets, each covering an interval of 100. Although this covers a relatively small portion of the overall range of citation values, this is the densest portion of the distribution and is thus deserving of closer scrutiny. A trend can be observed in the relative performance of certain models across the majority of the buckets, producing the ranking from better to worse: LSTM, MLP, linear regression, decision tree. The random forest is less consistent in its relative ranking across the buckets but typically performs worse than the LSTM and better than the decision tree. The k-nearest neighbors regression and the baseline always perform the worst among all the models but are not consistent in which performs better among each other. These results demonstrate a degree of consistency in the relative performance of the models across the densest portion of the distribution of citation values. To verify that this holds true for larger citation values where the distribution is much sparser, we plot the results for the range of 0-50,000 citations (Figure \ref{fig:cite_large}) broken into 5 buckets, each covering an interval of 10,000. With higher citation values, there is less stability in the order of the ranking. In particular, the error of the weighted delta baseline drops significantly, accounting for its low mean error in Table \ref{tab:citation_performance}. However across both the small and large buckets, the LSTM achieves the lowest error in all but one bucket, demonstrating excellent performance across a wide range of citation values to be predicted.

\begin{figure*} [!htp]
	\centering
	\includegraphics[width=1\textwidth]{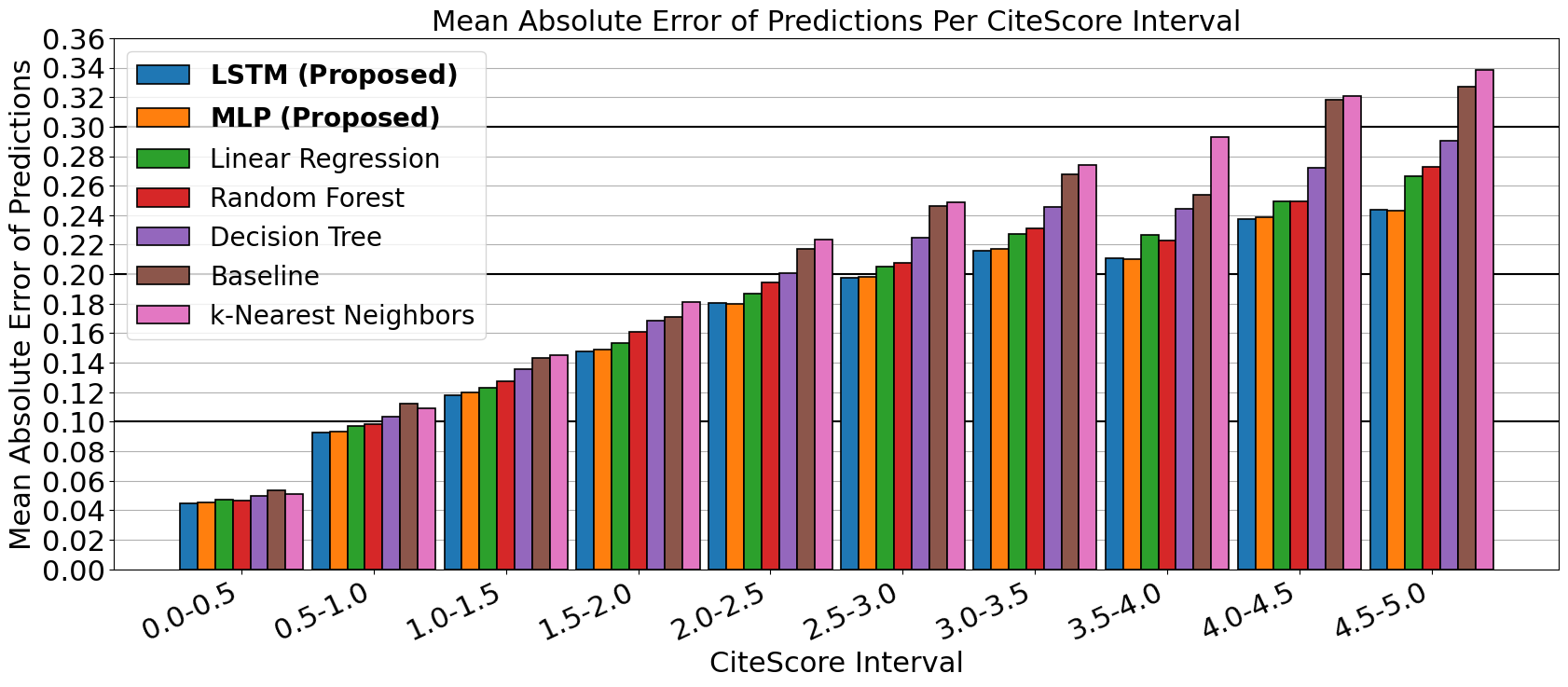}
	\caption{Mean absolute error of the selected model configurations for the CiteScore prediction task. Test samples with a ground truth CiteScore value in the range of 0-5 have been grouped into 10 buckets.} \label{fig:citescore_small}
\end{figure*}

\begin{figure*} [!htp]
	\centering
	\includegraphics[width=1\textwidth]{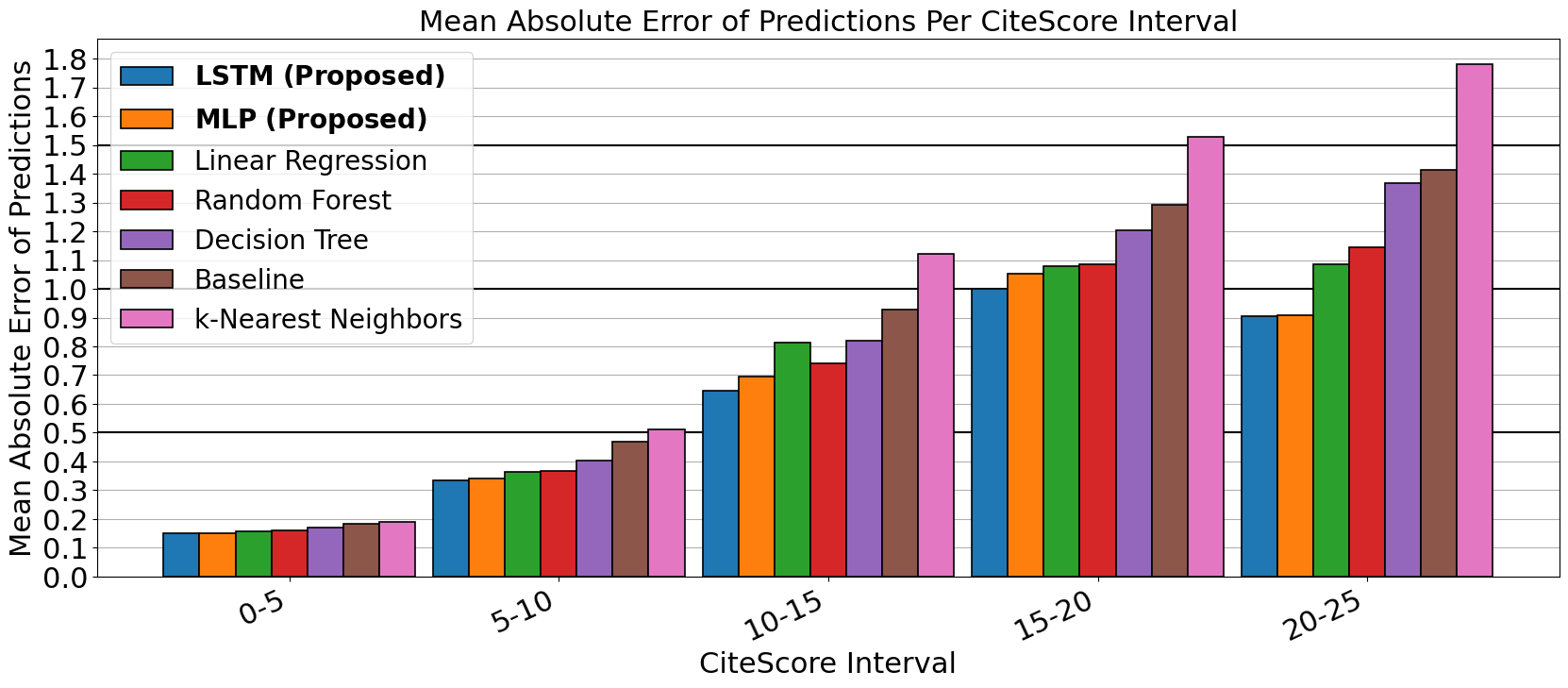}
	\caption{Mean absolute error of the selected model configurations for the CiteScore prediction task. Test samples with a ground truth CiteScore value in the range of 0-25 have been grouped into 5 buckets.} \label{fig:citescore_large}
\end{figure*}

\vspace{1cm}

The same experiment has been applied for the CiteScore prediction task covering CiteScore values in the range of 0-5 (Figure \ref{fig:citescore_small}) using 10 buckets and for values in the range of 0-25 (Figure \ref{fig:citescore_large}) using 5 buckets. We have limited the CiteScore values to this range due to a lack of sufficient test data for buckets covering larger CiteScore values. Few journals have CiteScore values above 25 and the available samples are further reduced by the 10\% test data split and the partitioning into buckets. As a result, buckets for larger values would have fewer than 20 samples each and could not be considered an accurate reflection of performance.

As with the citations prediction task, the same relative ranking of models can be observed across the majority of the buckets in these plots. For this prediction task, the random forest shows a greater stability in its relative ranking, generally performing worse than linear regression and better than the decision tree. It is interesting to note that for the CiteScore predictions, the LSTM achieves a much smaller improvement over the MLP than in the citations prediction task. We hypothesize that this is due to the predictive task being restricted to only the publications within the four year CiteScore window. Unlike in the task of predicting all citations received during the next year, longer sequences of historical data are more limited in their pertinence to the publications for which predictions are to be made in the CiteScore prediction task. This reduces the benefits gained from the use of an LSTM over an MLP in this task.

%prediction of publications is difficult and prediction of citations on very recent documents are difficult tasks. this is likely due to inherent unpredictability in the former case and a lack of historical information in the later case

\section{Analysis of the Selected LSTM Model} \label{sec:lstm_analysis}

Of the investigated models, the LSTM provides the best predictive performance in our experimental comparisons. To better understand the quality of the model, we perform a deeper analysis of its predictive results throughout this section and provide some discussion around its performance and the significance of its predictions.

\subsection{Justification for the Use of a Model}

We begin by examining the question of whether the quality of the predictive model is sufficient to justify its usage in practice. One might argue that the performance metrics of journals are slow-moving values and that there is little need to employ a complex predictive model in such a setting. To examine the degree to which the citation and CiteScore metrics of journals vary from one year to the next, we can refer back to the errors associated with the persistence baseline heuristic used in our experimental comparisons. Recall from Formula \ref{eq:persistence} that this heuristic works under the assumption that the value to be predicted remains similar from one year to the next and thus repeats the last known value as the prediction. The errors associated with this heuristic (as seen in the first row of Table \ref{tab:citation_performance} and \ref{tab:citescore_performance}) therefore reflect information about the expected change in the citation or CiteScore values from one year to the next.

To demonstrate the improvement achieved through the use of the LSTM model, we examine the percentage reduction in error of the MAE and MAPE for the LSTM in comparison to the corresponding errors of the persistence baseline. For the citations prediction task, the LSTM offers a 42.1\% reduction in the MAE and a 24.1\% reduction in the MAPE. For the CiteScore prediction task, the LSTM offers a 22.9\% reduction in the MAE and an 18.3\% reduction in the MAPE. These are significant reductions in error which justify the usage of a model in this setting. When critical decisions depend on the best information available, the benefit of a well-configured model over a heuristic based on the assumption of relative stability is clear.

\subsection{Further Analysis of the Predictive Results}

\begin{figure*} [!htp]
	\centering
	\includegraphics[width=1\textwidth]{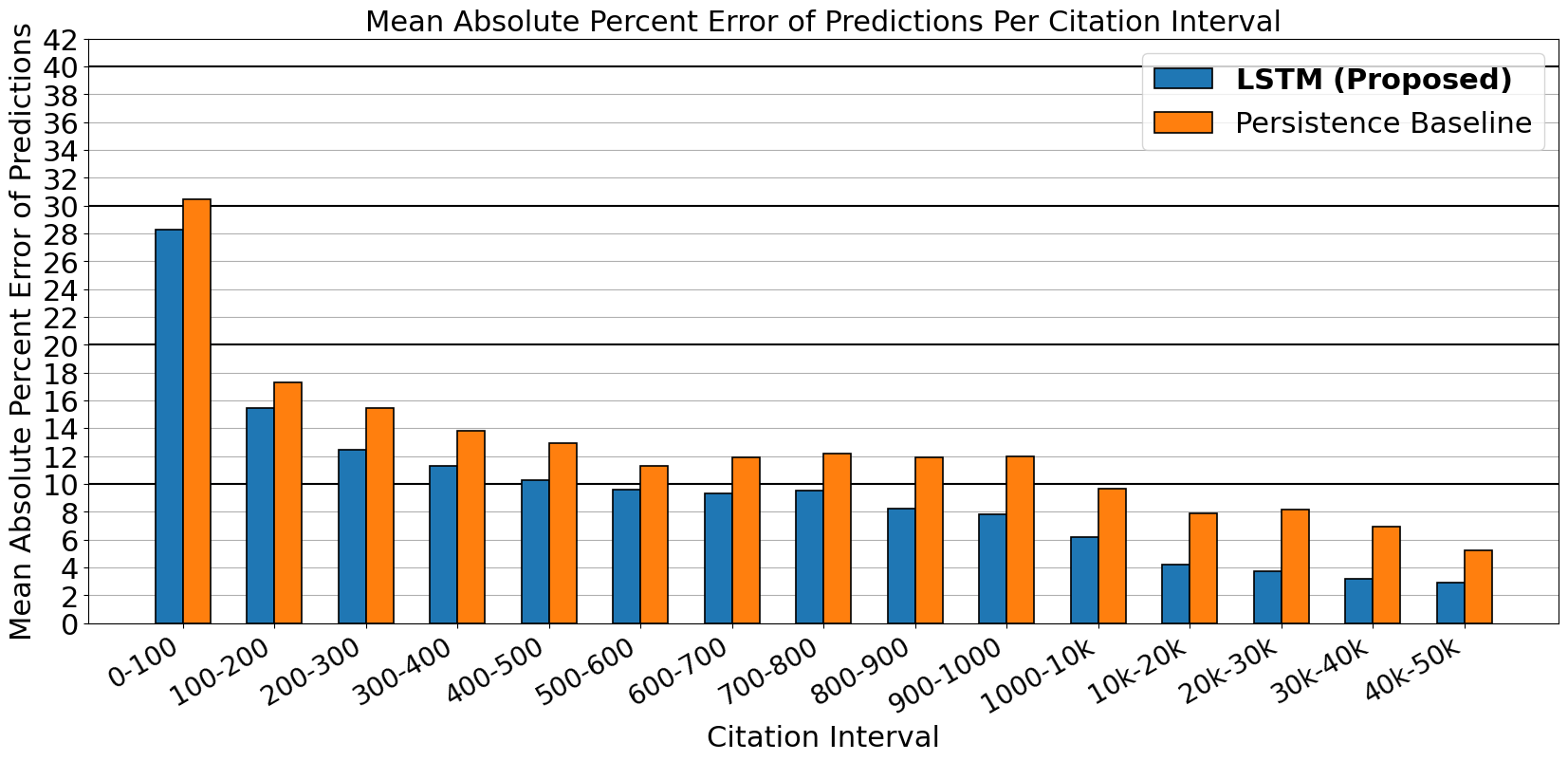}
	\caption{Mean absolute percent error of the LSTM and the persistence baseline for the citations prediction task. Test samples with a ground truth citation value in the range of 0-1,000 have been grouped into the first 10 buckets and test samples with a citation value in the range of 1,000 to 50,000 have been grouped into the final 5 buckets.} \label{fig:cite_percentage}
\end{figure*}

\begin{figure*} [!htp]
	\centering
	\includegraphics[width=1\textwidth]{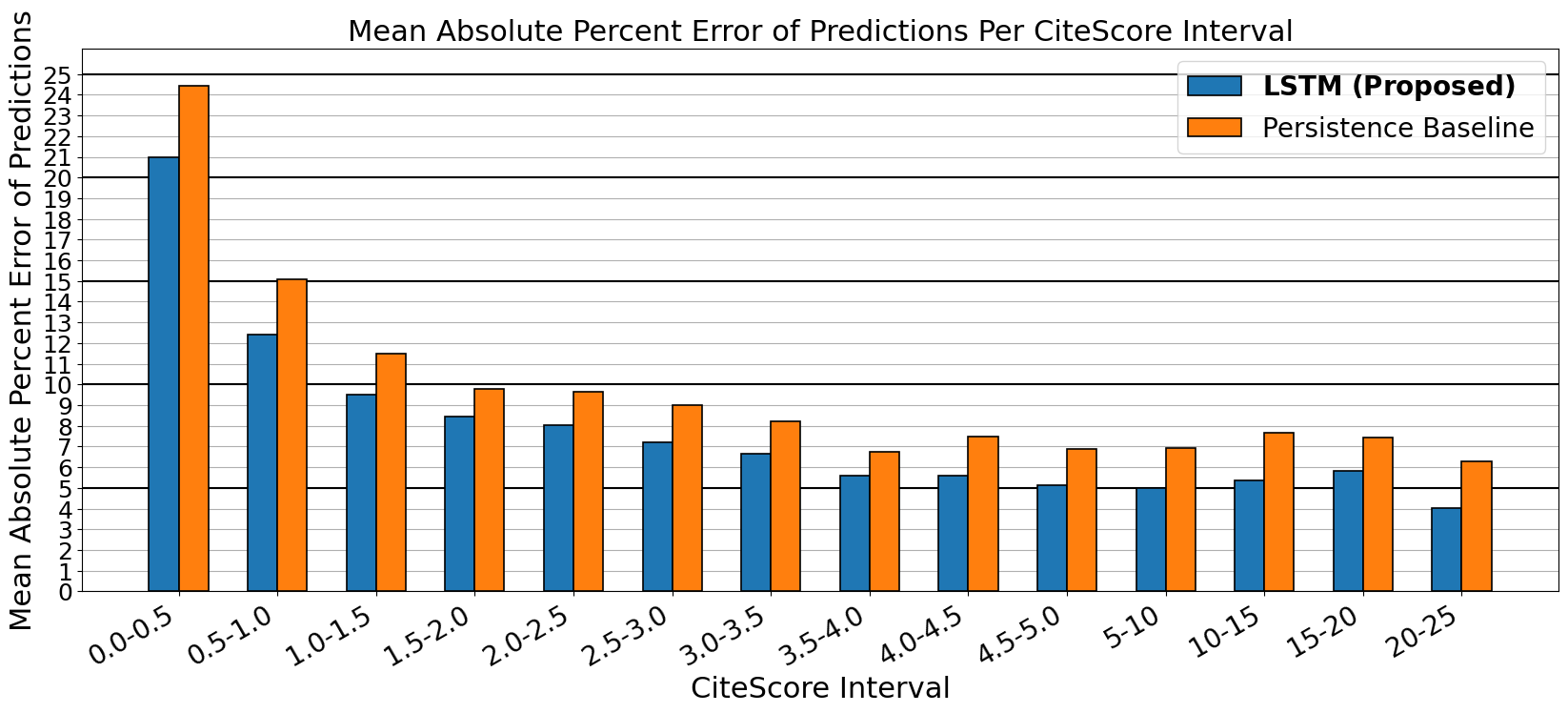}
	\caption{Mean absolute percent error of the LSTM and the persistence baseline for the CiteScore prediction task. Test samples with a ground truth CiteScore value in the range of 0-5 have been grouped into 10 buckets and test samples with a CiteScore value in the range of 5 to 25 have been grouped into the final 4 buckets.} \label{fig:citescore_percentage}
\end{figure*}

Next, we turn to a more detailed analysis of the predictive results provided by the LSTM model. Figures \ref{fig:cite_small} - \ref{fig:citescore_large} help to understand the distribution of the MAE across the range of values to be predicted. However, the difference in the scale of the values grouped into each bucket has implications on the quality of the predictions that may not be immediately apparent. To better observe this, it is useful to examine similar graphs of model performance shown in terms of the MAPE. We show this in Figure \ref{fig:cite_percentage} for the citations prediction task and Figure \ref{fig:citescore_percentage} for the CiteScore prediction task. We include the persistence baseline in these graphs for further comparison between our configured model and the na\"{i}ve approach.

It is interesting to note that for both predictive tasks, there is a decreasing trend in the MAPE for both the baseline and the LSTM as the bucket intervals increase. The observation of this trend in the persistence baseline suggests that journals with higher amounts of citations or higher CiteScores tend to have a greater stability in these values from year to year when interpreting the degree of change proportionally to the original value. Yet, the proportional improvement in performance of the LSTM over the persistence baseline also increases as the bucket intervals increase. In particular, there is roughly a 50\% reduction in error when using the LSTM for the largest citations buckets in comparison to working under the assumption that the number of citations remains the same between years. As a complex model drawing on multiple features over past years of journal performance, the LSTM is better suited to take advantage of the increased stability in these journals and deliver greater improvements in its predictions. 

For both the persistence baseline and the LSTM, the predictions in the first bucket have a high percentage of error. The high percent error in the persistence baseline indicates that the citation counts and CiteScores are much more volatile from one year to the next for journals with low values in these metrics when considering the proportionality of the change to the original value. The high percent error for the LSTM shows that this volatility is also more difficult to model. This is most likely due to the greater impact of noise on the small values to be predicted. For instance, a journal that typically receives a low number of annual citations may occasionally publish a paper that garners a large amount of attention, leading to a sudden spike in annual citations. The difference between the typical annual citations of the journal and those received during the first year of the spike may be a significant proportion of the base value, leading to a large percentage of error in the predictions. More sophisticated models that draw on publication-level data may be able to mitigate the effects of such phenomena on the quality of their predictions. However, it is important to keep in mind that despite the high MAPE value, the MAE of the citations and CiteScore predictions are already quite low.

\section{Conclusions} \label{sec:conclusion}

We have laid out two predictive tasks that are of use for projecting the future performance of academic journals, namely prediction of citations received during the next calendar year and prediction of the CiteScore assigned for the next calendar year. To perform these tasks using neural networks, we have collected training data via the Scopus APIs to create a dataset containing historical data on journal citations and CiteScore values, among other relevant data. We have designed experiments to perform feature selection and model configuration for MLP and LSTM models to accomplish the predictive tasks. Through comparisons with other classical machine learning models and heuristic prediction baselines, we have shown improved predictive power, particularly in the case of the LSTM, over these alternative methods. Through the use of diverse measures of error and a close examination of performance across the distributions of target citations and CiteScore values, we have demonstrated consistency in the quality of results from our predictive models.

There are a number of natural extensions to this work. The inclusion of additional input features may be able to provide improved performance from the predictive models. We limited ourselves to features available or computable from the Elsevier APIs. However, other sources of data such as altmetrics may provide a valuable complement to these features. If available, the use of features at a finer temporal granularity, such as monthly rather than yearly, may also provide a means to improve predictive abilities. Finally, while we have focused on two particular predictive tasks, the same methods can be adapted to design models for other related metrics such as Journal Impact Factor or to consider projection farther into the future to predict values multiple years in advance.

\section*{Acknowledgments}
We thank the Scopus Team and Elsevier for providing us with information about the Scopus database and guidance on the use of their APIs. These valuable interactions have helped us to ensure the creation of a high-quality dataset for this work.

\bibliography{Bibliography}

\end{document}